\documentclass[tradiabstract]{aa}

\usepackage{txfonts}
\usepackage{graphicx}
\usepackage{float}
\usepackage{natbib}
\usepackage[english]{babel}
\usepackage{color}
\usepackage{hyperref}
\usepackage[flushleft]{threeparttable}

\begin{document}

\title{The impact of clustering and angular resolution on far-infrared and millimeter continuum observations}

\author{Matthieu B\'ethermin\inst{1,2} \and  Hao-Yi Wu\inst{3,4} \and Guilaine Lagache\inst{1} \and Iary Davidzon\inst{1} \and Nicolas Ponthieu\inst{5} \and Morgane Cousin\inst{1} \and Lingyu Wang\inst{6,7} \and Olivier Dor\'e\inst{3,4} \and Emanuele Daddi\inst{8} \and Andrea Lapi\inst{9,10,11}}

\institute{Aix Marseille Univ, CNRS, LAM, Laboratoire d'Astrophysique de Marseille, Marseille, France, \email{matthieu.bethermin@lam.fr}
\and European Southern Observatory, Karl-Schwarzschild-Str. 2, 85748 Garching, Germany 
\and California Institute of Technology, MC 367-17, Pasadena, CA 91125, USA
\and Jet Propulsion Laboratory, California Institute of Technology, 4800 Oak Grove Drive, Pasadena, CA 91109, USA
\and Univ. Grenoble Alpes, CNRS, IPAG, F-38000 Grenoble, France
\and SRON Netherlands Institute for Space Research, Landleven 12, 9747 AD, Groningen, The Netherlands
\and Kapteyn Astronomical Institute, University of Groningen, Postbus 800, 9700 AV, Groningen, The Netherlands
\and CEA Saclay, Laboratoire AIM-CNRS-Universit\'e Paris Diderot, Irfu/SAp, Orme des Merisiers, F-91191 Gif-sur-Yvette, France
\and SISSA, Via Bonomea 265, 34136 Trieste, Italy
\and INAF-Osservatorio Astronomico di Trieste, via Tiepolo 11, 34131 Trieste, Italy 
\and INFN-Sezione di Trieste, via Valerio 2, 34127 Trieste, Italy}

\date{Received ??? / Accepted ???}

\abstract{Follow-up observations at high-angular resolution of bright submillimeter galaxies selected from deep extragalactic surveys have shown that the single-dish sources are comprised of a blend of several galaxies. Consequently, number counts derived from low- and high-angular-resolution observations are in tension. This demonstrates the importance of resolution effects at these wavelengths and the need for realistic simulations to explore them. We built a new 2\,deg$^2$ simulation of the extragalactic sky from the far-infrared to the submillimeter. It is based on an updated version of the 2SFM (two star-formation modes) galaxy evolution model.  Using global galaxy properties generated by this model, we used an abundance-matching technique to populate a dark-matter lightcone and thus simulate the clustering. We produced maps from this simulation and extracted the sources, and we show that the limited angular resolution of single-dish instruments has a strong impact on (sub)millimeter continuum observations. Taking into account these resolution effects, we are reproducing a large set of observables, as number counts and their evolution with redshift and cosmic infrared background power spectra. Our simulation consistently describes the number counts from single-dish telescopes and interferometers. In particular, at 350 and 500\,$\mu$m, we find that the number counts measured by \textit{Herschel} between 5 and 50\,mJy are biased towards high values by a factor $\sim$2, and that the redshift distributions are biased towards low redshifts. We also show that the clustering has an important impact on the \textit{Herschel} pixel histogram used to derive number counts from P(D) analysis. We find that the brightest galaxy in the beam of a 500\,$\mu$m \textit{Herschel} source contributes on average to only $\sim$60\,\% of the \textit{Herschel} flux density, but that this number will rise to $\sim$95\,\% for future millimeter surveys on 30 meter-class telescopes (e.g., NIKA2 at IRAM). Finally, we show that the large number density of red \textit{Herschel} sources found in observations but not in models might be an observational artifact caused by the combination of noise, resolution effects, and the steepness of color- and flux density distributions. Our simulation, called SIDES (Simulated Infrared Dusty Extragalactic Sky), is available at \url{http://cesam.lam.fr/sides}.} 

\keywords{Galaxies: statistics -- Galaxies: evolution -- Galaxies: star formation -- Galaxies: high-redshift -- Infrared: galaxies -- Submillimeter: galaxies}

\titlerunning{The impact of clustering and angular resolution on far-infrared and millimeter observations}

\authorrunning{B\'ethermin et al.}

\maketitle

\section{Introduction}

The star formation history (SFH) in the Universe is one of the key constraints to understand the evolution of galaxies. The combination of various tracers (H$_\alpha$, far UV, far infrared and millimeter) was successfully used in the last 20 years to measure the star formation rate density (SFRD) up to very high redshift ($z \sim 8$, see \citealt{Madau2014} for a review). At $z\ge$2\,-\,3, building complete spectroscopic samples becomes very challenging and continuum emission is mainly used to derive star formation rates (SFR). Consequently, the prime tracer of recent star formation is the redshifted far-UV emission from young stars. However, even at early epochs, massive galaxies have already formed a large amount of dust and UV light is thus absorbed \citep[e.g.,][]{Takeuchi2005,Heinis2014}. Two main approaches can then be used to derive the intrinsic SFR: correct the UV absorption using the UV spectral slope as a proxy of attenuation \citep[e.g.,][]{Calzetti2000} or directly detect the reprocessed UV light emitted by dust in the far-infrared and millimeter \citep{Kennicutt1998}.

Far-infrared and submillimeter observations are challenging because of the limited angular resolution of the instruments. The deepest observations of the most modern single-dish instruments are limited by the confusion, that is, the blending of sources in the same beam of the instrument \citep[e.g.,][]{Dole2003}. Only the brightest galaxies emerge from the confusion and can be extracted individually from far-infrared and submillimeter maps. However, because of the large beam of the single-dish instruments, their measured flux density can be contaminated by their fainter neighbors. Indeed, follow-up observations of the brightest 850\,$\mu$m sources at high-resolution with ALMA  revealed that a large fraction are multiple sources \citep[e.g.,][]{Karim2013,Hodge2013}. Because of this, the flux density distributions measured with single-dish instruments and interferometers such as ALMA \citep{Karim2013,Simpson2015} strongly disagree.

Also, the \textit{Herschel} space observatory \citep{Pilbratt2010} has a limited angular resolution and could be affected by similar effects. However, it is very difficult to verify because interferometric follow-up observations are not possible at the high frequencies of the \textit{Herschel} observations. Consequently, other approaches such as modeling must be used to explore possible biases induced by the angular resolution on \textit{Herschel} number counts. In particular, we have to understand why the number density of red \textit{Herschel} sources found in the extragalactic surveys is almost an order of magnitude higher than that predicted by the models (\citealt{Asboth2016}, see also \citealt{Dowell2014,Ivison2016}).

In addition to studies of bright sources above the confusion limit, various advanced techniques were developed to probe galaxy populations in the confusion such as the stacking method \citep[e.g.,][]{Dole2006,Marsden2009,Bethermin2010a,Viero2013b}, P(D) measurements \citep[e.g.,][]{Condon1974,Patanchon2009,Glenn2010}, or source extraction using position priors coming from shorter wavelengths \citep[e.g.,][]{Magnelli2009,Bethermin2010b,Roseboom2010,Hurley2017}. These methods can also be biased by the contamination of the measured flux by faint clustered sources.

Simulations were developed to test these possible biases \citep[e.g.,][]{Fernandez_conde2008}, but the clustering of infrared galaxies at high redshift was poorly constrained at that time. Important progress has been made recently. In particular, \textit{Planck} and \textit{Herschel} measured cosmic infrared background (CIB) anisotropies with an unprecedented precision \citep{Planck_CIB2011,Amblard2011,Planck_CIB2013,Viero2013}. Their modeling showed that the typical mass of the dark-matter halos hosting the bulk of the obscured star formation is almost constant and around 10$^{12}$\,M$_\odot$ up to z$\sim$3 \citep[e.g.,][]{Bethermin2012a,Bethermin2013,Viero2013,Planck_CIB2013,Wu2016}. In addition, clustering studies of bright high-redshift far-infrared and millimeter galaxies showed they are hosted by massive halos ($\sim 10^{13}$\,M$_\odot$, e.g., \citealt{Farrah2006,Weiss2009,Magliocchetti2014,Bethermin2014,Wilkinson2017}). These massive halos are strongly clustered. The impact of clustering on the extraction of sources from confusion-limited surveys might be stronger than predicted by pre-\textit{Planck} and \textit{Herschel} simulations, which assumed a weaker clustering.

It is thus timely to develop new simulations that are able to reproduce simultaneously the far-infrared and millimeter observations at various angular resolutions. These simulations must include clustering and take into account all the lessons learnt from \textit{Herschel} and ALMA. On the one hand, \citet{Hayward2013c} built a simulation based on abundance matching, but this analyzes only the galaxy populations selected at 850\,$\mu$m (see also \citealt{Munoz-Arancibia2015,Cowley2015}). On the other hand, \citet{Schreiber2016} built a simulation of the panchromatic properties of galaxies, but did not include a physical clustering model. Our new simulation combines the strengths of these two approaches and
accurately reproduces spectral and spatial properties of galaxies and CIB anisotropies. In this paper, we focus on the continuum properties of galaxies and the effect of angular resolution from 70\,$\mu$m to 1.2\,mm. In a future paper, we will introduce the (sub)millimeter line ([CII], [NII], [CI], CO...) properties of galaxies, discuss the perspectives for (sub)millimeter intensity mapping and test methods of line deblending.

Our simulation is based on the Bolshoi-Planck simulation \citep{Klypin2016,Rodriguez-Puebla2016}, from which a lightcone covering 2\,deg$^2$ was produced. We populate the dark-matter halos using an abundance-matching technique \citep[e.g.,][]{ValeOstriker2004}. The luminous properties of the galaxies are derived using an updated version of the 2SFM (2 star-formation modes) model \citep{Sargent2012,Bethermin2012c,Bethermin2013}. This model is based on the observed evolution of the main sequence of star forming galaxies \citep[e.g.,][]{Noeske2007,Elbaz2007,Daddi2007}, that is, a SFR-M$_\star$ correlation evolving with redshift, and the observed evolution of the spectral energy distributions (SEDs) with redshift. In the new version of the model, we take into account the increase of dust temperature in main sequence galaxies recently measured from z=2 to z=4 \citep{Bethermin2015a}, extending the increase found from z=0 to z=2 by \citet{Magdis2012b}. We also include the latest calibration of the evolution of the main sequence \citep{Schreiber2015}.

In Sect.\,\ref{sect:prescriptions}, we present the ingredients of our simulation and discuss its limitations. We compare our results with observed number counts and discuss the effects of resolution in Sect.\,\ref{sect:counts}. We then discuss the redshift-dependent observables and the consequences on the obscured star formation history (Sect.\,\ref{sect:redshift}). We then show the significant impact of clustering on the pixel histogram of the \textit{Herschel} maps, also known as P(D), and check that our model correctly reproduces the CIB anisotropies measured by \textit{Herschel} and \textit{Planck} (Sect.\,\ref{sect:pdpk}). Finally, we discuss the existence of the red sources found by \textit{Herschel} surveys (Sect.\,\ref{sect:red}).

We assume a \citet{Planck2015_cosmo} cosmology and a \citet{Chabrier2003} initial mass function (IMF). The products of our simulation, called SIDES (Simulated Infrared Dusty Extragalactic Sky), are publicly available at \url{http://cesam.lam.fr/sides}.

\section{Ingredients of the simulation}

\label{sect:prescriptions}


This section describes the ingredients used to build our simulated sky, namely,
\begin{itemize}
\item the dark-matter lightcone, which is the starting point (Sect.\,\ref{sect:dm}),
\item the stellar mass function (Sect.\,\ref{sect:smf}),
\item the abundance-matching procedure used to populate the dark-matter halos with galaxies (Sect.\,\ref{sect:am}),
\item our recipe to split galaxies into a star forming and a passive population (Sect.\,\ref{sect:sffrac}),
\item our method to derive a SFR for each galaxy (Sect.\,\ref{sect:sfrprop}),
\item the assignment of SEDs to our simulated galaxies (Sect.\,\ref{sect:sed}),
\item the implementation of strong and weak lensing (Sect.\,\ref{sect:lensing}).
\end{itemize}
Finally, in Sect.\,\ref{sect:limitations}, we discuss the limitations of our simulation. We homogenized the cosmology used in the dark matter simulation and in the observed stellar mass functions. Our method is described in Appendix\,\ref{sect:homogenization}.

\subsection{Dark matter simulation and lightcone catalog}

\label{sect:dm}

We use the publicly available halo catalogs from the Bolshoi-Planck simulation \citep{Rodriguez-Puebla2016}\footnote{The catalogs are available at\\ \url{http://hipacc.ucsc.edu/Bolshoi/MergerTrees.html}}. The simulation has a volume of (250 $h^{-1}$ Mpc)$^3$, with a dark-matter particle mass of $1.5 \times 10^8 h^{-1}M_\odot$. The cosmological parameters are compatible with \cite{Planck2015_cosmo}:
$h = 0.678$,
$\sigma_8 = 0.823$,
$\Omega_\Lambda  = 0.693$,
$\Omega_M  = 0.307$,
$\Omega_b =0.048$,
$n_s = 0.96$.

Dark matter halos are identified by the phase-space halo finder Rockstar \citep{Behroozi2013Rockstar}. We use the halo mass M$_{200}$, which is defined by the radius within which the spherical overdensity is 200 times the critical density of the Universe.  We only use halos with mass above $10^{10}$\,M$_\odot$, which have more than 50 dark matter particles. We have explicitly verified that above $10^{10}\,M_\odot$, the halo mass function from our simulation agrees with the analytic halo mass function.

Using the simulation snapshots at different redshifts, we construct a lightcone catalog of 1.4 deg$\times$ 1.4 deg, $0<z<10$, corresponding to a comoving volume of 0.17 ${\rm Gpc^3}$, approximately three times the volume of the Bolshoi-Planck simulation. In a lightcone catalog, each object is at a cosmic distance that corresponds to the cosmic time that it emits light.  The simulation outputs are saved at discrete time steps, and we use snapshots approximately spaced by $\Delta z = 0.25$.  We have
explicitly checked that the structure in such a narrow redshift bin has negligible evolution and can be represented by the same snapshot.

To construct the lightcone, we replicate the box in all three dimensions, using the periodic boundary condition inherent in the simulations. Since our lightcone catalog has a pencil-beam geometry, we use a "slanted'' line of sight to reduce the repeated structure; that is, the line of sight is not parallel to any of the axes or diagonals of the box. Specifically, we first rotate the box by $10^\circ$ along the $y$-axis and another $10^\circ$ along the $z$-axis.  We then transform the cartesian coordinates into the equatorial coordinates, following the convention of {\tt astropy}. The distance of an object from the observer is converted into the cosmological redshift, and we add to the redshift the peculiar velocity along the line-of-sight. For more details about the constructions of the lightcone, we refer to \cite{Merson2013}.

\begin{figure*}
\centering
\includegraphics[width=1\columnwidth]{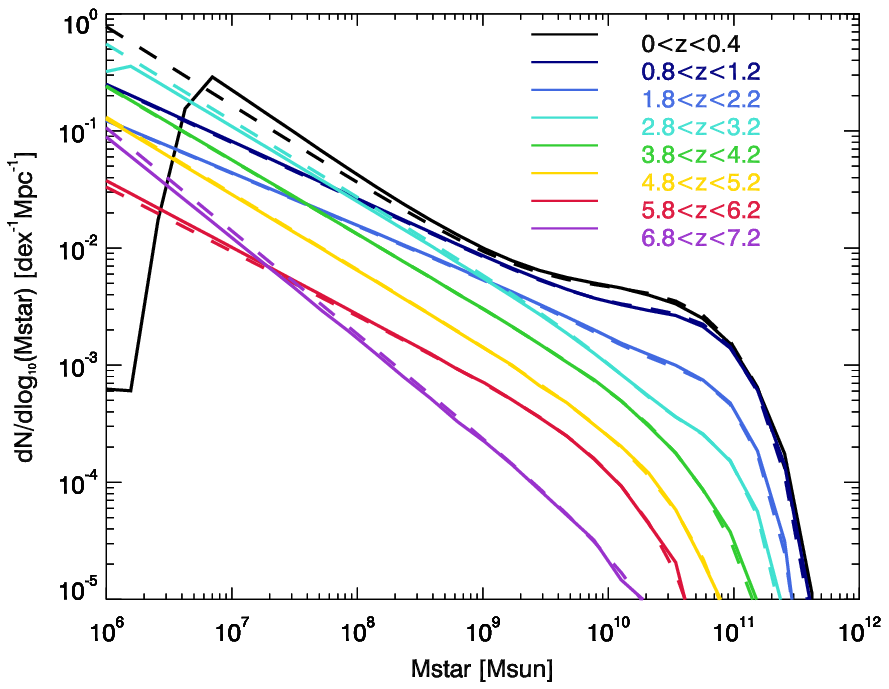}
\includegraphics[width=1\columnwidth]{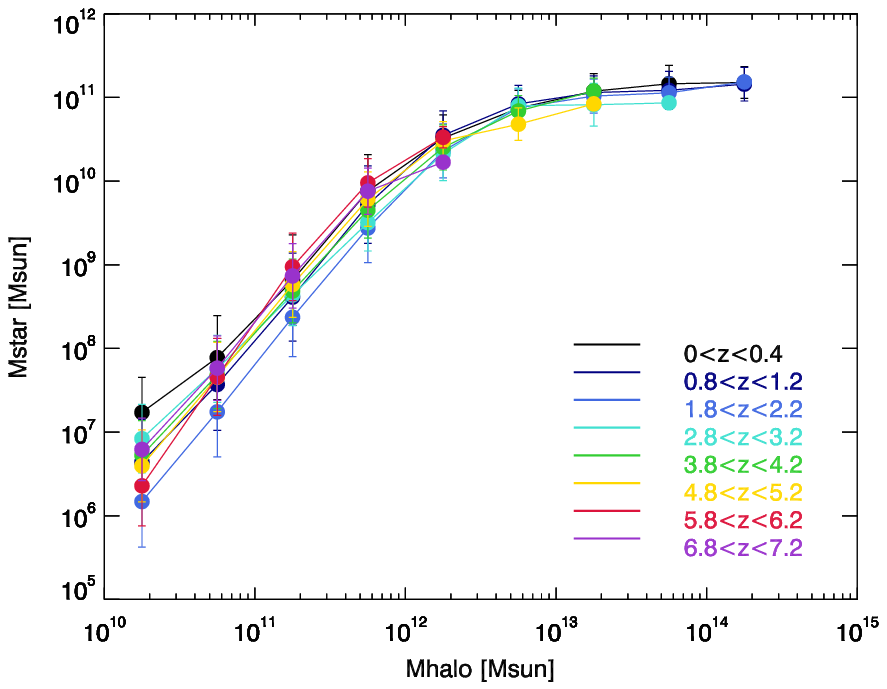}
\caption{\label{fig:SMHM}\textit{\bf Left:} Evolution of the stellar mass function used in our simulation with redshift. The construction of this stellar mass function is explained in Sect.\,\ref{sect:smf}. The solid line is the SMF computed from the catalog in a bin $\Delta z$=0.4 . The dashed line is the analytic SMF at the mean redshift of the bin. The cut at low mass (M$_\star <10^7$\,M$_\odot$) is due to the halo mass limit of the dark-matter simulation. \textit{\bf Right:} Relation between the halo mass and the stellar mass from our abundance-matching procedure (see Sect.\,\ref{sect:am}). The error bar indicates the 0.2\,dex scatter on this relation.}
\end{figure*}

\subsection{Stellar mass function}

\label{sect:smf}


In our simulation, the stellar mass function (SMF) is the starting point from which we generate all the properties of the galaxies. Similarly to the approach presented in \citet{Bernhard2014}, we assume that it can be described by a double Schechter function \citep[e.g.,][]{Baldry2012}:
\begin{equation}
        \phi(\rm M_\star) ~\rm d(M_\star)= e^{-\frac{\rm M_\star}{\rm \mathcal{M}^\star}}~ \left [\Phi_1^\star \left (\frac{\rm M_\star}{\rm \mathcal{M}^\star} \right)^{\alpha_1}+\Phi^{*}_2 \left (\frac{\rm M_\star}{\rm \mathcal{M}^{*}} \right )^{\alpha_2} \right ] \, \frac{\rm d(M_\star)}{\rm \mathcal{M}^\star},
\label{eq:smf}
\end{equation}
where $\mathcal{M}^\star$ is the characteristic mass of the knee of the SMF, $\Phi^{*}_1$ and $\Phi^{*}_2$ are the normalization of the two components, and ${\alpha_1}$ and ${\alpha_2}$ the power-law slopes at low mass. We use the same functional representation of the SMF at all redshifts to avoid discontinuities of its evolution with redshift.

The evolution with redshift of the parameters described above is based on the observations. We use the data points of \citet{Kelvin2014} in the GAMA field in the local Universe, \citet{Moutard2016b} from the VIPERS survey up to z=1.5, \citet{Davidzon2017} in the COSMOS field from z=1.5 to z=4, and \citet{Grazian2015} at z$>$4. \citet{Grazian2015} uses a simple Schechter function. At z$>$4, to ensure a smooth transition with smaller redshifts at which a double Schechter function is used, we fix $\Phi_1^\star$ to 0 and use the $\Phi$ and $\alpha$ of the single Schechter function for the second component. We connect the data points (taken at the center of the redshift bins of the authors) using a linear interpolation of each parameter (log($\mathcal{M}^\star$), $\Phi_1^\star$, log($\Phi_2^\star$), $\alpha_1$, $\alpha_2$) as a function of (1+z). We chose to use $\Phi_1^\star$ and log($\Phi_2^\star$) to avoid problems with the log where $\Phi_1^\star$ is fixed to zero and to avoid negative values at z$>7$, respectively. The stellar mass function of the galaxies used to generate our simulation is shown in Fig.\,\ref{fig:SMHM}. Below 10$^8$\,M$_\odot$, the number density at fixed mass no longer evolves monotonically with redshift. This unphysical behavior is caused by the uncertainties on the low-mass slope of the observed data we used. This is a limitation of our empirical approach. However, these low-mass sources have a small impact on our simulation, since they emit only 4\,\% of the infrared luminosity. Thus, we have chosen to keep these sources in the simulation, since they contribute to confusion noise.

\subsection{Abundance matching}

\label{sect:am}


To assign stellar mass to dark matter halos and subhalos, we perform subhalo abundance matching between the halo catalogs and the stellar mass functions described above. The basic idea of abundance matching is to assign higher stellar mass to more massive halos or subhalos, either monotonically or with some scatter, according to the number densities of the objects in the Universe \citep[e.g.,][]{ValeOstriker2004,Shankar2006,Behroozi2013,Moster2013}. In this work, instead of mass, we use the peak circular velocity $v_{\rm pk}$ of dark matter halos and subhalos to perform the abundance matching, since $v_{\rm pk}$ is known to be more tightly correlated with stellar mass \citep[e.g.,][]{Reddick2013}.

We assume that the stellar mass has an intrinsic scatter of 0.2 dex at a given $v_{\rm pk}$, which is required for the resulting galaxy catalog to reproduce the observed galaxy clustering \citep{Reddick2013}. The input SMF (Eq.\,\ref{eq:smf}) is deconvolved into a stellar mass function without the intrinsic scatter on the stellar versus halo mass relation; this deconvolved stellar mass function is then used to match the number density of halos monotonically.  We use the implementation by Y.-Y. Mao\footnote{The code is publicly available at\\ \url{https://bitbucket.org/yymao/abundancematching}}, and we refer the readers to \citet{Behroozi2010,Behroozi2013} and \citet{WuDore2016} for the detailed implementation. The left panel of Fig.\,\ref{fig:SMHM} demonstrates that the stellar mass functions resulted from this abundance-matching calculation (solid curves) recover the input stellar mass functions (dashed curves). There is a slight tension at low mass in some redshift bins caused by the evolution of the SMF inside a redshift bin. We also observe a sharp cut below 10$^7$\,M$_\odot$ for $0<z<0.4$, which is caused by the halo mass limit of the simulation. Since halo and stellar masses are correlated, this also implies a low-mass cut in the stellar mass function.

The right panel of Figure~\ref{fig:SMHM} shows the stellar mass--halo mass relation resulting from the abundance-matching calculation.  

\subsection{Fraction of star forming galaxies}

\begin{figure}
\centering
\includegraphics{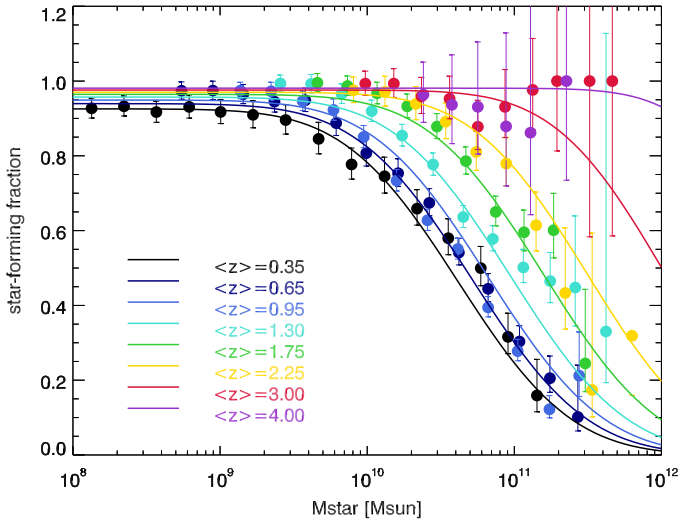}
\caption{\label{fig:sffrac} Fraction of star-forming galaxies versus stellar mass at various redshifts. The data points are from \citet{Davidzon2017}. The solid line is a fit by the parametric form described in Sect.\,\ref{sect:sffrac}.}
\end{figure}

\label{sect:sffrac}

We will draw randomly galaxy properties from their stellar mass and redshift using the prescriptions of the 2SFM formalism \citep{Sargent2012,Bethermin2012c}, which applies only to star forming galaxies. First, we have to estimate the probability of a galaxy at a given $M_\star$ and redshift to be star forming. Accordingly, we split the galaxies in our simulation in two populations: passive galaxies, which have a negligible star formation, and star forming galaxies. We used the observed evolution of the star forming fraction by \citet{Davidzon2017} to derive this fraction. In this work, the authors classified the galaxies as star forming or not using their position in the (NUV-r) versus (r-K) color diagram \citep{Arnouts2013,Ilbert2013}. We fit their results with the following parametric form (see also Fig.\,\ref{fig:sffrac}):
\begin{equation}
\label{eq:sffrac}
f_{\rm SF}(M_\star, z) = (1 - f_{\rm Q,0}(z)) \, \frac{1 - \textrm{erf} \Big[ \frac{\textrm{log}_{10}(M_\star) - \textrm{log}_{10}\big(M_t(z)\big)}{\sigma_{\rm SF}(z)}\Big]}{2 },
\end{equation} 
where $f_{\rm Q,0}(z)$ is the fraction of passive galaxies at low mass ($M_\star << M_t(z)$). The fraction is higher at low redshift, where a significant fraction of low-mass galaxies in dense environments are passive. In Eq.\,\ref{eq:sffrac}, $M_t(z)$ is the stellar mass of the transition between passive and star forming galaxies, and $\sigma_{\rm SF}(z)$ the width of this transition. These three quantities evolve with redshift. Their evolution is parametrized in the following way:
\begin{equation}
f_{\rm Q,0}(z) = f_{\rm Q,0,z=0} (1+z)^\gamma
,\end{equation}
\begin{equation}
\textrm{log}_{10}(M_t)(z) = \textrm{log}_{10}(M_{t,z=0}) + \alpha_1 z + \alpha_2 z^2
,\end{equation}
\begin{equation}
\sigma_{\rm SF}(z) = \sigma_{\textrm{SF},z=0}+ \beta_1 z + \beta_2 z^2.
\end{equation}
This parametric form provides an excellent fit of the measurements (reduced $\chi^2$ of 0.82). The best fit parameters are $ f_{\rm Q,0,z=0} = 0.1017$, $\textrm{log}_{10}(M_{t,z=0}) = 10.53$, $\sigma_{\textrm{SF},z=0} = 0.8488$, $\alpha_1 = 0.2232$, $\alpha_2 = 0.0913$, $\beta_1 = 0.0418$, $\beta_2 = -0.0159$, and $\gamma = -1.039$.

This approach neglects environmental effects, since it depends only on the stellar mass and redshift. Our simulation is optimized for field galaxies, cosmic infrared background, and intensity mapping studies. Cosmic infrared background and intensity mapping are dominated by central galaxies and are thus not severely affected by these effects \citep[e.g.,][]{Bethermin2013}. The limitations implied by this simplification are discussed in Sect.\,\ref{sect:limitations}.

\subsection{Star-forming properties}

\label{sect:sfrprop}

We assume that only galaxies classified as star-forming have far-infrared and millimeter outputs. In passive galaxies, some residual emission of cirrus heated by the old stellar populations has been observed. However, at a given stellar mass, these galaxies usually have infrared luminosities lower by at least one order of magnitude than galaxies on the main sequence \citep[e.g.][]{Viero2013b,Amblard2014,Man2016,Gobat2017,Gobat2017b}. Neglecting their infrared outputs is thus a fair assumption.

The SFR of star forming galaxies is derived using the 2SFM formalism \citep{Sargent2012,Bethermin2012c}. The first step is to compute the mean SFR of sources from the measured evolution of the main sequence of star forming galaxies, written afterwards $\textrm{SFR}_{\rm MS}$. \citet{Schreiber2015} measured the evolution of this main sequence up to $z=4$ and proposed the following parametric description:
\begin{equation}
\label{eq:msdesc}
\begin{split}
\textrm{log}_{10}\left(\frac{\textrm{SFR}_{MS}}{M_\odot/yr}\right) = \textrm{log}_{10}\left(\frac{M_\star}{10^9\,M_\odot}\right) - m_0 + a_0 \textrm{log}_{10}(1+z) \\ 
- a_1 \left[\textrm{max}\left( 0,\textrm{log}_{10}\left(\frac{M_\star}{10^9\,M_\odot}\right) - m_1 - a_2 \textrm{log}_{10}(1+z) \right) \right]^2,\end{split}
\end{equation}
with $m_0 = 0.5$, $a_0 = 1.5$, $a_1 = 0.3$, $m_1 = 0.36$, $a_2 = 2.5$. In \citet{Bethermin2012c}, we assumed a simple power law for the main sequence at a given redshift. In addition, at z$>$2.5, there was no evolution of sSFR, that is, $\textrm{SFR}/M_\star$, with z at fixed $M_\star$. In this updated version (Eq.\,\ref{eq:msdesc}), the SFR decreases sharply at high $M_\star$ and sSFR continues to evolve at higher redshift. This rising sSFR was already discussed in \citet{Bethermin2013}, since it reproduces the CIB anisotropies better. The \citet{Schreiber2015} formula is fitted on observations at z$>$0.5 and sSFR is too high at lower redshift. To correct for this offset, we applied a $0.1 \times \frac{0.5-z}{0.5-0.22}$\,dex offset to the \citet{Schreiber2015} formula at z$<$0.5. The detailed explanations are provided in Appendix\,\ref{sect:lowzcorr}.

Star forming galaxies are not all on the main sequence. In this paper, a starburst is defined as a positive outlier of the main sequence \citep[e.g.,][]{Elbaz2011,Sargent2012}. Following \citet{Bethermin2012c}, the fraction of starburst does not vary with stellar mass; it grows linearly with redshift from 1.5\% at $z=0$ to 3\% at z=1 and stays flat at higher redshift. We randomly drew a main sequence or a starburst galaxy using this probability.

The main sequence is of course not a perfect correlation and it has a non-negligible scatter. We followed a procedure similar to \citet{Bethermin2012c} to distribute the galaxies around the main sequence. We randomly drew the SFR of each source using a log-normal distribution in agreement with the observational results \citep[e.g.,][]{Rodighiero2011}. Following \citet{Sargent2012}, the \citet{Bethermin2012c} model assumed a width of 0.15 and 0.2\,dex for main sequence galaxies and starbursts, respectively. However, more recent measurements by \citet{Schreiber2015} and \citet{Ilbert2015} found a slightly higher width of 0.3\,dex (see also \citealt{Sun2016}). We thus use this updated value in our simulation. The distribution of main sequence galaxies is centered on 0.87\,SFR$_{\rm MS}$ and 5.3\,SFR$_{\rm MS}$ for the starbursts \citep{Schreiber2015}. Since a log-normal distribution centered on 1 has a mean value above unity, the center of main sequence is set to 0.87\,SFR$_{\rm MS}$ in order to have the correct mean SFR (see \citealt{Schreiber2015} and \citealt{Ilbert2015} for more explanations).

Using follow-up observations of submillimeter sources detected by single-dish telescopes with interferometers, \citet{Karim2013} showed that the brightest of these sources have multiple components (see also \citealt{Simpson2015}). They found that the bright end of the number counts at 850\,$\mu$m were significantly overestimated and that none of the single components have a SFR significantly above 1000\,M$_\sun$/yr. The SFR distribution of 870\,$\mu$m-selected galaxies measured by \citet{Da_Cunha2015} drops strongly above 1000\,M$_\odot$/yr. A rapid drop of the number density at SFR$\gtrapprox$1000\,$M_\odot$ was also found by high-resolution radio observations \citep{Barger2014,Barger2017}. For simplicity, we implemented a sharp SFR limit at 1000\,M$_\sun$/yr. The SFR of each galaxy is redrawn until it is lower than this limit. Consequently, the sSFR distribution of the most massive galaxy populations is truncated at high sSFR. Wide surveys found some rare sources with a higher SFR suggesting this is not a sharp limit \citep[e.g.,][]{Riechers2013,Tan2014,Ma2016}. However, using a sharp limit rather than an exponential cut of the sSFR distribution is a reasonable assumption considering the small size of our field. The impact of this SFR limit on the number counts is discussed in Sect.\,\ref{sect:sfrlim}. So far, the physical origin of this SFR cut is not totally clear. These objects could be Eddington-limited starbursts limited by the radiative pressure \citep[e.g.,][]{Thompson2005}. This could also be explained by the weaker boost of star formation induced by mergers in gas-rich systems \citep[e.g.,][]{Fensch2017}. 

\subsection{Spectral energy distributions and continuum fluxes}

\label{sect:sed}

\begin{figure}
\centering
\includegraphics[width=1\columnwidth]{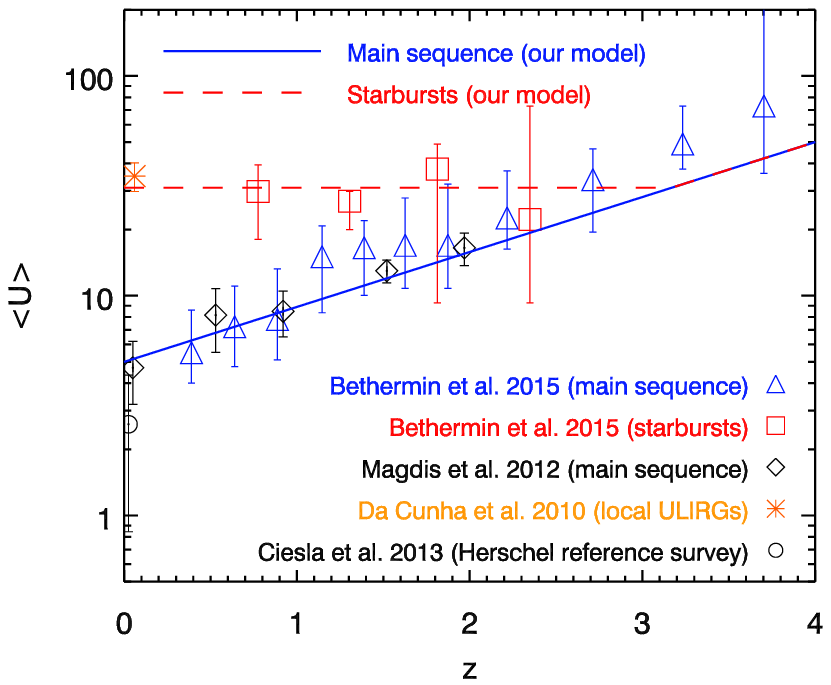}
\caption{\label{fig:sed}Evolution of the mean intensity of the radiation field $\langle$U$\rangle$ (see \citealt{Magdis2012b}) that is used in the current updated model. The evolution of main sequence galaxies and starbursts are plotted in blue (solid line) and red (dashed line), respectively. Also shown are the measurement of \citet[][triangles for main sequence galaxies and squares for starbursts]{Bethermin2015a}, \citet[][diamonds]{Magdis2012b}, \citet[][asterisk]{DaCunha2010}, and \citet[][circle]{Ciesla2014}. This Figure is adapted from \citet{Bethermin2015b}.}
\end{figure}

We then assigned a SED to each of our sources to derive their flux densities in a large set of instrument filters from their total infrared luminosity ($L_{\rm IR}$). $L_{\rm IR}$ is directly derived from SFR using the \citet{Kennicutt1998} conversion factor (1.0$\times$10$^{-10}$\,M$_\odot$/yr/L$_\odot$ after converting to \citealt{Chabrier2003} IMF). We use \citet{Magdis2012b} SED library. The shape of the SEDs depends on the galaxy type (main sequence or starburst) and on the $\langle$U$\rangle$ parameter, that is, the mean intensity of the radiation field. This parameter is strongly correlated with the dust temperature \citep[e.g.,][]{Dale2002}. It evolves with redshift for main sequence galaxies (see Fig.\,\ref{fig:sed}). In \citet{Bethermin2012c}, we had no data above $z=2$ and we assumed a flattening at $z>2$, since it provides a better agreement with the observed submillimeter number counts. Two new observational inputs motivated us to update the evolution of $\langle$U$\rangle$ in our simulation. In \citet{Bethermin2015a}, we measured that $\langle$U$\rangle$ continues to rise at z$>$2 using a  stacking analysis. In addition, \citet{Karim2013} also showed that submillimeter number counts were overestimated because of blending effects (see the discussion in Sect.\,\ref{sect:multi}). The previous measurements favored a scenario with no evolution of $\langle$U$\rangle$ at z$>$2, because it was producing colder SEDs and consequently higher submillimeter counts. This is no longer true with the new number counts that are observed to be lower.

For main sequence galaxies, we used $\langle U_{\rm MS} \rangle (z=0) = 5$ as in \citet{Bethermin2012c}. An evolution in $(1+z)^\alpha$ does not fit very well the observational data from \citet{Bethermin2015a} with an overly sharp decrease with decreasing redshift at $z<0.5$. This artificially low $\langle U \rangle$ at low redshift is responsible of an excess of the bright number counts at 160\,$\mu$m. We thus used another parametric form, which fits better the observational data:
\begin{equation}
\textrm{log}_{10}\big[\langle U_{\rm MS} \rangle(z)\big] = \textrm{log}_{10}\big[\langle U_{\rm MS} \rangle(z=0)\big] + \alpha_{\langle U \rangle} z
,\end{equation}
with $\alpha_{\langle U \rangle} = 0.25$. Following \citet{Bethermin2015a}, we use a constant $\langle U_{\rm SB}\rangle$ = 31. However, at z$>3$, this would lead to starbursts colder than main sequence galaxies. This behavior could be considered as unphysical and we thus assumed $\langle U_{\rm MS} \rangle = \langle U_{\rm SB} \rangle$ at higher redshift. At z$>4$, we have no constraints on an evolution of $U_{\rm MS}$. Extrapolating this behavior up to z$\sim$10 would imply unphysically high values. We thus assume a plateau in the high redshift regime (z$>$4). In addition to this mean evolution, we also included a 0.2\,dex scatter on $\langle U \rangle$ following \citet{Magdis2012b}.

\subsection{Magnification by lensing}

\label{sect:lensing}

Gravitational lensing can have a non-negligible impact on the bright submillimeter number counts, because of their steepness \citep{Negrello2007,Negrello2010,Bethermin2011,Lapi2011,Bethermin2012c,Lapi2012,Vieira2013,Wardlow2013,Negrello2017}. At 350 and 500\,$\mu$m, this effect is maximal around 100\,mJy, where $\sim$20\,\% of the sources are lensed. Our simulation of a 2\,deg$^2$ field contains only six sources brighter than this threshold. The lensing has thus a relatively weak effect on the total number counts; however, it has a non-negligible impact on the number of bright red sources (see Sect.\,\ref{sect:red}), since the fraction of lensed sources is higher at high redshift. It is therefore important to consider lensing.\\

For each source of our simulation, we randomly drew the magnification $\mu$. The determination of the magnification does not include any spatial information (see Sect.\,\ref{sect:limitations} for a discussion about this approximation). For the strong lensing ($\mu > 2$), we used the probability distribution of \citet{Hezaveh2011} used also in \citet{Bethermin2012b}, which depends only on the redshift. We also included a simplified weak lensing model for the other sources. We randomly drew their magnification from a Gaussian, whose width and mean value are derived from \citet[][their Fig.\,1 and 2]{Hilbert2007}.

\subsection{Limits of our simulation}

\label{sect:limitations}

Our simulation is based on the observed evolution of star forming galaxies and aims to accurately reproduce current observations of the far-infrared and (sub)millimeter Universe. However, the current version of this simulation has several limitations, which should be kept in mind while comparing it with observations. Our simulation is based on a single 2\,deg$^2$ field. Since it is based on a dark-matter simulation, it is thus affected by cosmic variance beyond simple Poisson fluctuations and can contain under- or overdensities at some specific redshift.

Our abundance-matching procedure assumes that the stellar mass of a galaxy is associated with $v_{\rm pk}$ (proxy of the potential well of dark matter halos or subhalos), with some scatter. During our abundance-matching procedure, we implicitly assume that main halos and subhalos follow the same relation. In addition, the probability of a galaxy to be passive at a given z depends only
on its stellar mass. Our simulation thus neglects the environmental quenching observed in the most massive halos (M$_{\rm halo} > 10^{14}$\,M$_\odot$) at z$<$1 \citep[e.g.,][]{Peng2010}. There are only 26 such halos in our simulation. Moreover the contribution of these massive structures to the star formation density is small \citep{Popesso2015a}. This approximation should thus only be a problem if the simulation is used to study low-redshift overdensities.

Our description of the lensing  depends only on the redshift and ignores the position of foreground sources. This treatment is thus inconsistent with the large-scale structures of our simulation. Overdensities of low-z galaxy populations are associated with massive halos, which can strongly magnify high-redshift sources \citep[e.g.,][]{Wang2011,Welikala2016}. A full consistent treatment of lensing is beyond the scope of this paper and we thus decided to have a purely probabilistic treatment of lensing magnification. The impact of this simplification should be small on most of the statistics, but spatial correlations between bright-lensed sources and their neighbors could be significantly affected.

Finally, at z$>$4, our simulation relies on extrapolations of relations calibrated at lower redshift. The SEDs used in our simulation evolve only up to z=4. At higher redshift, we assume no evolution due to the lack of constraints. Potentially, SEDs could become even warmer because of the effect of CMB \citep{DaCunha2013b,Zhang2016}. However, the temperature of our SEDs is higher than the dust temperature assumed in these studies ($\sim$ 40\,K versus $\sim$15\,K). The CMB effect should thus be much smaller than estimated in these studies, but it might be non-negligible for the highest-redshift objects of our simulation. The evolution of sSFR in our simulation is based on the \citet{Schreiber2015} relation, which is derived from z$<4$ data. The scatter on the main sequence is also assumed to be constant with mass and redshift, since there is currently no evidence of the contrary. Finally, the evolution of the parameters of the stellar mass function are extrapolated at z$>$6. The predictions of our simulation at z$>$4 should thus be taken with caution.

Our simulation currently contains only the far-infrared and millimeter observables and we thus assumed in Sect.\,\ref{sect:sed} that L$_{\rm IR}$ traces the total star formation. However, in low-mass and high-redshift galaxies, the fraction of UV photons escaping the galaxies can be non-negligible. The impact of neglecting unobscured star formation on the infrared observables was discussed extensively in \citet{Bernhard2014}. They showed that the scatter of the infrared excess (IRX = L$_{\rm IR}$ / L$_{\rm UV}$) has a negligible impact on infrared observables. In contrast, the increasing IRX with increasing M$_\star$ implies that low-mass objects have a smaller fraction of their UV reprocessed by dust and the faint-end slope of the number counts should be slightly steeper if we include the UV \citep{Bethermin2012c}. The impact of these lower number counts on the confusion noise is small ($<$5\,\%), since the shot noise is proportional to $dN/dS S^2 dS$, where S is the flux density and $dN/dS$ are the number counts.

\section{Number counts and multiplicity of sources detected by single-dish instruments}

\label{sect:counts}

In this section, we demonstrate that our simulation is able to reproduce the observed number counts, when the effects of angular resolution on source extraction are properly taken into account. We also discuss in particular the multiplicity of \textit{Herschel}/SPIRE and NIKA2 sources and the bias caused by clustering on stacking measurements.

\subsection{Simulating the observational process}

\label{sect:simobs}

\citet{Karim2013} showed that the 850\,$\mu$m sources found by single-dish telescopes are often blends of several sources. The same phenomenon could also impact other single-dish observations and especially \textit{Herschel}. We thus compare the measurements with both the intrinsic number counts from our simulated catalog and number counts extracted from simulated maps.

We built \textit{Herschel} simulated maps from our simulated catalog. We used Gaussian beams with full widths at half maximum (FWHM) of 5.5, 6.5, 11, 18.2, 24.9, and 36.3\,arcsec at 70, 100, 160, 250, 350, and 500\,$\mu$m, respectively, corresponding to the measured size of the \textit{Herschel} beams. We did not include any instrumental noise, since we are only interested in the effect of angular resolution. The faint sources in the simulated map are responsible for the confusion noise. We measured a confusion noise of 6.0, 6.5, and 6.0\,mJy at 250, 350, and 500\,$\mu$m, respectively. This is compatible at 2\,$\sigma$ with the measurements of \citet{Nguyen2010}, who found 5.8$\pm$0.3, 6.3$\pm$0.4, and 6.8$\pm$0.4, respectively.

We extracted the sources from \textit{Herschel} maps using FASTPHOT \citep{Bethermin2012b}. This routine uses source positions from another wavelength as a prior to deblend their flux. 
A large fraction of \textit{Herschel} catalogs were produced using  the position of 24\,$\mu$m sources as a prior \citep[e.g.,][]{Roseboom2010,Berta2011,Bethermin2012c,Magnelli2013}. 

Photometry routines using positional priors are not converging when too many sources are located in the same beam because of degeneracies. We thus kept only the brightest 24\,$\mu$m sources in a 0.5\,FWHM of radius in our list of prior position. Finally, even if catalogs extracted using position priors are not affected by flux boosting, they are still affected by the Eddington bias. This bias appears when a steep distribution is convolved by measurement uncertainties (see \citealt{Bethermin2012b}). We estimated the correction factor following \citet{Bethermin2012c}. We start from the flux distribution measured in the map. We then add a random Gaussian noise to each flux and compare the flux distribution before and after adding this noise. The photometric noise is estimated using the standard deviation of the residual map.

Contrary to $\lambda \leq $500\,$\mu$m, we cannot use 24\,$\mu$m priors to extract the sources at 850\,$\mu$m and 1.2\,mm, since the 24\,$\mu$m is no longer probing the dust emission ($>$8\,$\mu$m rest-frame) at the typical redshift of the sources detected at these wavelengths (see Sect.\,\ref{sect:redshift}). Thus, we extracted blindly the $>$5\,$\sigma$ peaks in our simulated map. This task is relatively easy, since we have no instrumental noise in our simulations. We then measured the flux density of the detected sources using FASTPHOT and deboosted the fluxes following \citet{Geach2017}. 

Single-dish observations were performed with various angular resolutions. We chose to use the resolution of the James Clerk Maxwell Telescope (JCMT, 15-meter diameter). This choice was guided by the fact that the most recent single-dish surveys at 850\,$\mu$m and 1.1\,mm / 1.2\,mm were performed with this telescope. Usually, ground-based (sub)millimeter maps are convolved by the Gaussian of the size of the beam before extracting the sources. This technique is optimal to extract point sources in noise-limited maps, but it increases the confusion and blending problems. The convolved map is usually called beam-smoothed map. We thus produced beam-smoothed simulated maps, with an effective FWHM after convolution of 21 and 26\,arcsec at 850\,$\mu$m and 1.1\,mm, respectively. We used the beam at 1.1\,mm instead of 1.2\,mm, since the most accurate number counts were measured at this wavelength with the AzTEC camera \citep{Scott2012}.

\begin{figure*}
\centering
\includegraphics{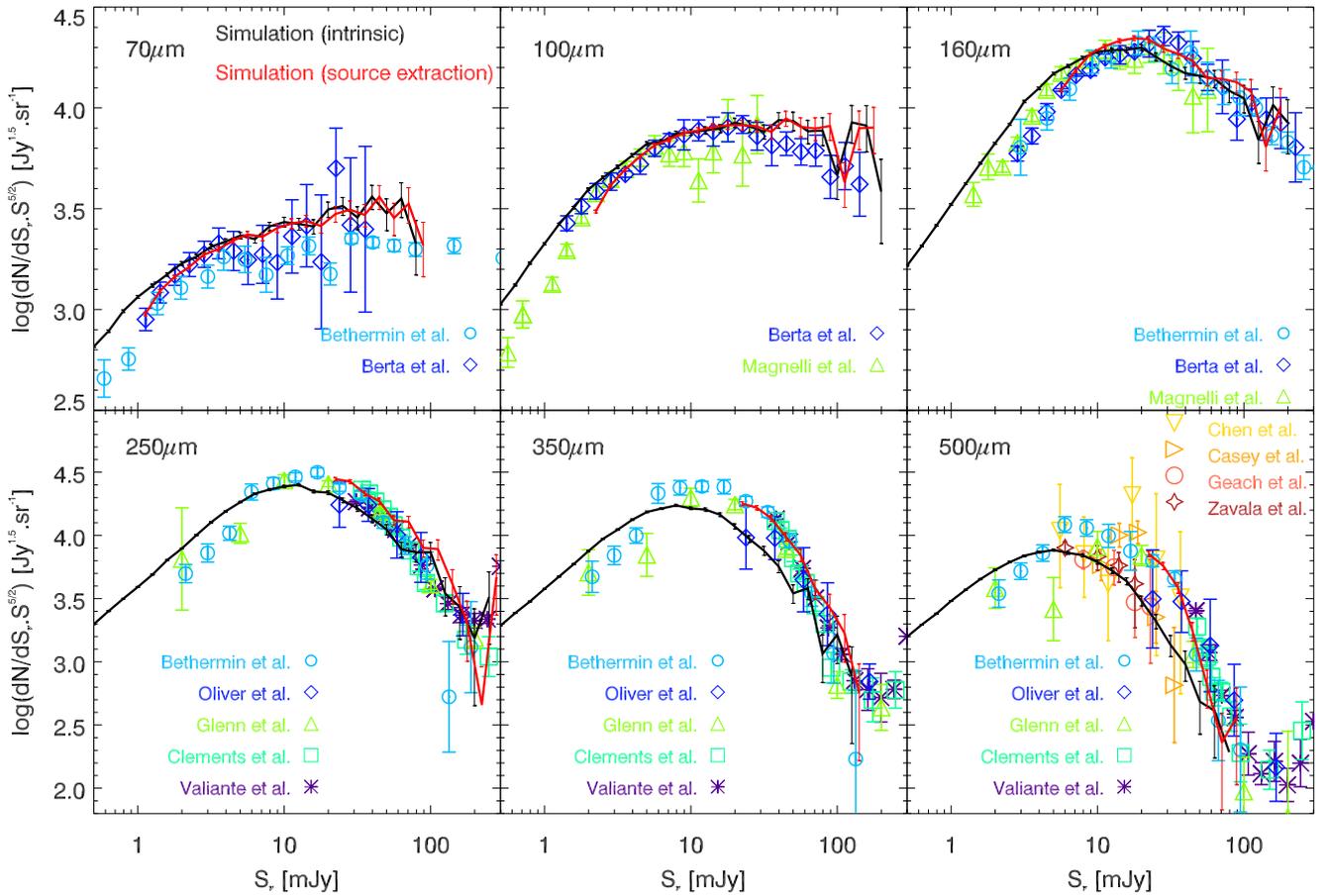}
\caption{\label{fig:herschel_counts} Differential number counts from 70\,$\mu$m to 500\,$\mu$m. The counts are multiplied by S$_\nu^{2.5}$ to reduce the dynamic range of the plot and to highlight the plateau at high flux densities where the Euclidian approximation is valid \citep[e.g.,][]{Planck_eucl}. The solid black line is the prediction from our simulated catalog. The red line is derived from the extraction of sources in simulated \textit{Herschel} maps using a method similar to that used by \citet{Bethermin2012c}. The error bars on the prediction of the simulation are derived assuming Poisson statistics. The source extraction procedure is limited by confusion and only bright sources can be extracted reliably. From 70\,$\mu$m to 160\,$\mu$m, the light blue circles, the dark blue diamonds, and the green triangles are the \textit{Spizter} counts of \citet{Bethermin2010a}, \textit{Herschel}/PEP counts of \citet{Berta2011}, and \textit{Herschel}/GOODS counts of \citet{Magnelli2013}, respectively. From 250\,$\mu$m to 500\,$\mu$m, the light blue circles, the dark blue diamonds, the green triangles, turquoise squares, and purple asterisks are the \textit{Herschel}/SPIRE measurements of \citet{Bethermin2012b}, \citet{Oliver2010b}, \citet{Glenn2010}, \citet{Clements2010}, and \citet{Valiante2016}, respectively. The SCUBA2 measurements at 450\,$\mu$m of \citet{Chen2013}, \citet{Casey2013}, \citet{Geach2013}, and \citet{Zavala2017} are shown using gold down-facing triangles, orange right-facing triangles, red open circles, and brown stars, respectively.}
\end{figure*}

\begin{figure*}
\centering
\includegraphics{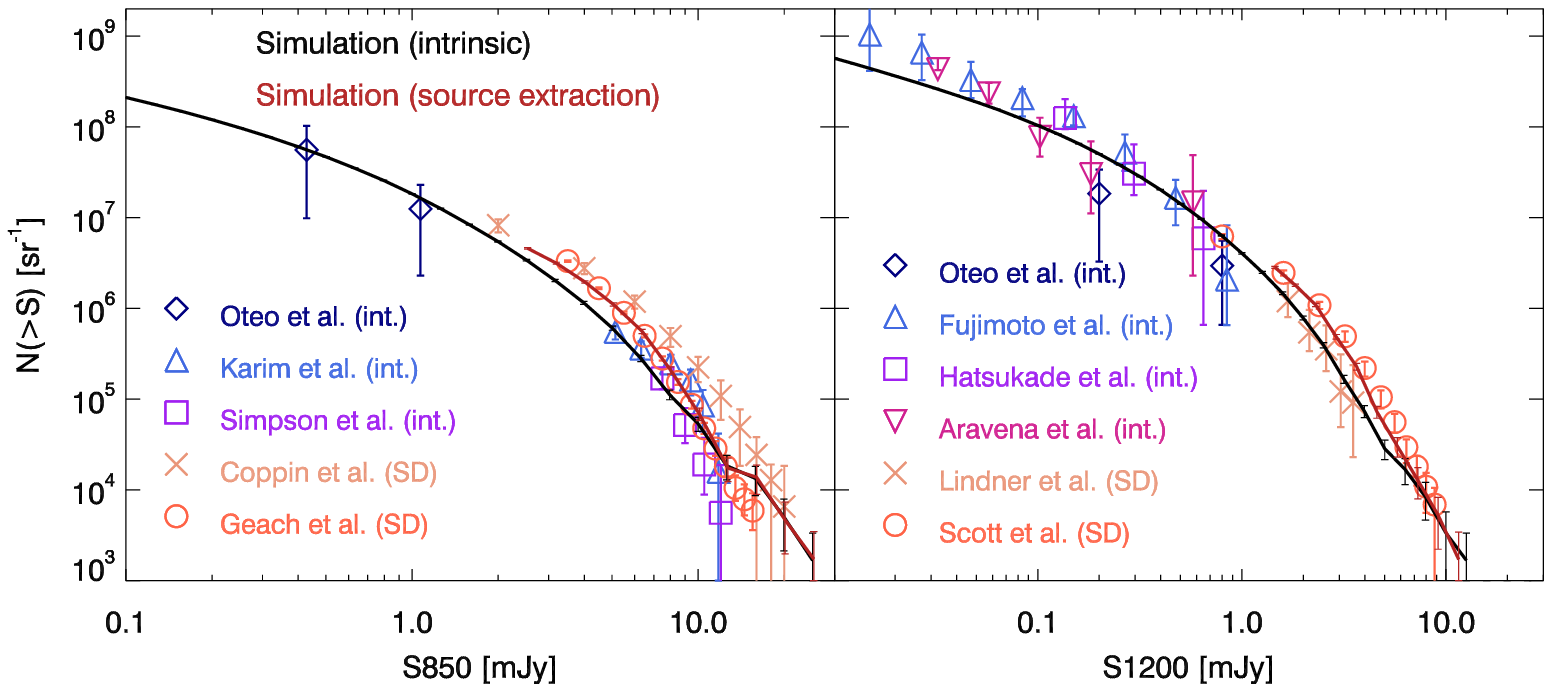}
\caption{\label{fig:submmcounts} Integral number counts at 850\,$\mu$m and 1.2\,mm. The black solid line is the intrinsic counts from our simulated catalog. The red solid line is the result of source extraction from the simulated map of single-dish instrument (see Sect.\,\ref{sect:submm_counts}). The number counts extracted from high-resolution interferometric data (int.) are colored in blue and purple\citep{Oteo2016,Karim2013,Simpson2015,Fujimoto2016,Hatsukade2013,Aravena2016}, while we use orange for single-dish (SD) results \citep{Coppin2006,Geach2017,Lindner2011,Scott2012}.}
\end{figure*}

\subsection{\textit{Spitzer} and \textit{Herschel} number counts}

\label{sect:herschel_counts}

The comparison between our simulation and the observed number counts is presented in Fig.\,\ref{fig:herschel_counts}. The intrinsic number counts in our simulated catalog (black solid lines) agree well with the data overall. However, there are some tensions at some specific wavelengths and flux regimes. The number counts at 70\,$\mu$m in our simulation (both intrinsic and extracted from the simulated maps) are 2\,$\sigma$ high at the bright end compared with \citet{Bethermin2010a} \textit{Spitzer} measurements, but agree at 1\,$\sigma$ with the \textit{Hercshel}/PACS measurements of \citet{Berta2011}. The intrinsic faint-end slope ($<$2\,mJy) of the PACS number counts (70, 100, and 160\,$\mu$m) is less steep in our simulation than in the observations, but the number counts recovered after a source extraction in our simulated map (red solid lines, Sect.\,\ref{sect:simobs}) agree with the observations. Jin et al. (in prep.) also found that the published PACS number counts are underestimated using advanced source extraction techniques.

Below 5\,mJy, the intrinsic \textit{Herschel}/SPIRE number counts (250, 350, and 500\,$\mu$m) are 2\,$\sigma$ higher than the constraints derived by stacking by \citet{Bethermin2012b} and by P(D) analysis by \citet{Glenn2010}. These constraints come essentially from the GOODS fields, which are deep but small and thus strongly affected by the cosmic variance. For instance, only the S$<$5\,mJy data points of \citet{Bethermin2012b} come from GOODS-N. The S$>$5\,mJy data points are dominated by COSMOS, which probes a much larger volume than the GOODS fields, and agree well with our simulation at 250\,$\mu$m. In addition, the pixel histograms of the COSMOS maps, that is,\ P(D), which is very sensitive to the number of faint sources (see Sect.\,\ref{sect:pdpk}), agree well with our simulation.

The main disagreement between intrinsic and measured number counts is located between 5\,mJy and 50\,mJy at 350\,$\mu$m and 500\,$\mu$m, where the simulation is a factor of 2 below the measurements. In contrast, the number counts extracted from the simulated maps (red solid line) agree well with the observations. The resolution has thus a strong impact on the bright \textit{Herschel}/SPIRE number counts and models should thus be compared with observations only after having simulated these resolution effects. Consequently, models adjusted directly on the observed number counts potentially overestimate the number of bright dusty star forming galaxies. 

The SCUBA2 camera observed deep fields at 450\,$\mu$m with a 8\,arcsec angular resolution \citep{Chen2013,Casey2013,Geach2013,Zavala2017}. In Fig.\,\ref{fig:herschel_counts}, these data points are shown using yellow, orange, and brown colors. We did not attempt to correct for the slightly different wavelength, since the 450\,$\mu$m/500\,$\mu$m color varies strongly with redshift. The latest data points of \citet{Zavala2017} agree very well with the intrinsic number counts in our simulation. This is not surprising, because the much better resolution of SCUBA2 compared with SPIRE limits the effect of resolution on the number counts. Our simulation also well agree with \citet{Chen2013} and \citet{Geach2013}. \citet{Casey2013} measurements have a 3\,$\sigma$ excess between 10 and 20\,mJy and disagree with both the previously quoted measurements and our simulation.

\subsection{Ground-based (sub)millimeter number counts}

\label{sect:submm_counts}

Contrary to number counts at $\lambda \leq $500\,$\mu$m, number counts at 850 and 1.2\,mm were measured with both interferometers and single-dish telescopes. \citet{Karim2013} (see also \citealt{Simpson2015}) showed that number counts derived using low- and high-angular-resolution data are inconsistent. These wavelengths are thus essential to test the ability of our simulation to
consistently describe these resolution effects. The comparison between our simulation and the observed number counts at 850\,$\mu$m and 1.2\,mm is presented in Fig.\,\ref{fig:submmcounts}. In order to homogenize these data taken at heterogeneous wavelengths, we applied a multiplicative factor of 0.8 to the 1.1\,mm data to convert them at 1.2\,mm and a factor of 1.07 to 870\,$\mu$m data to convert them at 850\,$\mu$m. These factors are derived using our main sequence SED template at z=2 and are only weakly redshift dependent.

At 850\,$\mu$m, our model agrees well with sub-mJy number counts extracted by \citet{Oteo2016} using ALMA calibration observations. Above 1\,mJy, we have access to two types of constraints: single-dish measurements (in orange, \citealt{Coppin2006,Geach2017}) and interferometric follow-up of these bright single-dish sources (blue and purple, \citealt{Karim2013,Simpson2015}). As explained in \citet{Karim2013}, the number counts derived from the interferometric follow-up of bright sources are lower than the number counts extracted directly from single-dish data, because the flux density of some single-dish sources is coming from several galaxies. Our intrinsic number counts agree perfectly with the interferometric data. The number counts extracted from the simulated map agree well with \citet{Geach2017}, but are slightly lower than \citet{Coppin2006}.

At 1.2\,$\mu$m, our intrinsic number counts are in good agreement with the deep blank ALMA fields \citep{Hatsukade2013,Fujimoto2016,Aravena2016,Oteo2016}). The number counts extracted from the simulated single-dish maps agree perfectly with \citet{Scott2012} and are 1\,$\sigma$ higher than \citet{Lindner2011}. \citet{Scott2012} used a mix of ASTE and JCMT data. The angular resolution is thus similar to that used in our simulated map. \citet{Lindner2011} data were taken with the IRAM 30-meter telescope and have thus an angular resolution two times higher, which explains why these measurements are significantly below the number counts extracted from the simulated map. In contrast, they agree well with the intrinsic number counts of the simulated catalog.

We thus managed to reproduce simultaneously the interferometric and single-dish number counts at 850\,$\mu$m and 1.2\,mm together with those from \textit{Herschel}. This reconciles the observations at low- and high-angular-resolution and highlights the importance of taking into account both the clustering and resolution effects in the modeling of the evolution of dusty galaxies.

\begin{figure*}
\centering
\includegraphics{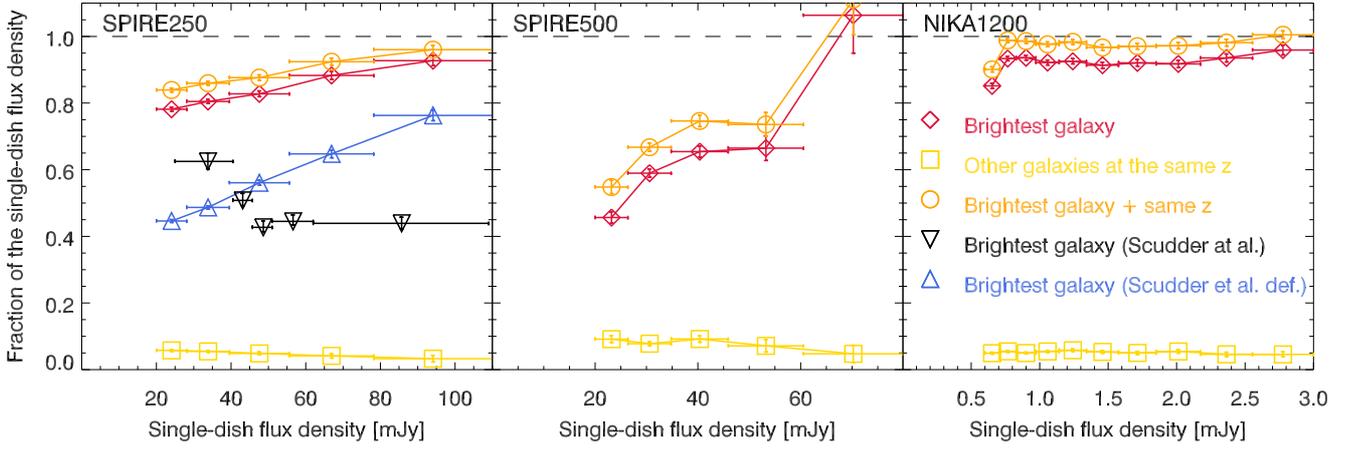}
\caption{\label{fig:multi} Average fraction of the flux density emitted by the brightest galaxy in the beam as function of the flux density measured in our simulated map with a limited angular resolution (red diamonds). We present our results for \textit{Herschel}/SPIRE at 250\,$\mu$m (left) and 500\,$\mu$m (middle), and for NIKA2 at 1.2\,mm (right). The gold squares show the average contribution of the galaxies physically related ($|\Delta z| < 0.01$) to the brightest galaxy in the beam. The orange open circles are the sum of the flux density fraction from the brightest galaxy and the other galaxies at the same redshift. The black downward-facing triangles is the flux density fraction from the brightest galaxy measured by \citet{Scudder2016}. They used a different definition from ours and divided the flux density of the brightest galaxy by the total of the flux density of all the galaxies in a 1\,FWHM radius. The blue upward-facing triangles are the results from our simulation assuming their definition (see Sect.\,\ref{sect:multi}).}  
\end{figure*}

\subsection{Multiplicity of single-dish sources}
\label{sect:multi}

As we shown in Sect.\,\ref{sect:herschel_counts} and \ref{sect:submm_counts}, number counts at $\lambda \geq 350$\,$\mu$m derived from single-dish observations are severely affected by the limited angular resolution of the instruments. We thus expect that the flux density of bright single-dish sources is emitted by several galaxies. This phenomenon has been well studied at 850\,$\mu$m from both an observational and theoretical point of view \citep{Karim2013,Hayward2013,Hodge2013,Cowley2015}. In contrast, it is much less explored for \textit{Herschel} sources, because of the difficulty in observing with interferometers from the ground below 850\,$\mu$m. In this paper, we present the results of our simulation of the \textit{Herschel} sources and predictions for the new NIKA2 camera at IRAM \citep{Monfardini2011}.

For each single-dish source extracted from the simulated map with the method described in Sect.\,\ref{sect:simobs}, we searched in our simulated catalog for the brightest galaxy in the beam. We used a search radius of 0.5\,FWHM, since the brightest galaxy is usually close to the center of the single-dish source ($<$0.15\,FWHM on average for \textit{Herschel} and NIKA2 data) and we want to avoid selecting a galaxy contributing to another close single-dish source. We then computed the ratio between the flux density of this brightest galaxy in our simulated catalog and the flux density of the single-dish source measured in our simulated map. In Fig.\,\ref{fig:multi}, we show the average ratio as a function of the measured single-dish flux density.

We also estimated the fraction of the flux density emitted by other galaxies at a similar redshift as the brightest galaxy. We chose to define a redshift as similar if $|\Delta z| < 0.01$. This value was determined using the histogram of the difference between the redshift of the brightest source and the other sources in the beam. This histogram has a very sharp peak around $\Delta z = 0$ with a FWHM of 0.0072, 0.0064, 0.0047 for SPIRE 250\,$\mu$m, SPIRE 500\,$\mu$m, and NIKA2 1.2\,mm, respectively. Our $|\Delta z| < 0.01$ criterion thus corresponds to at least 3\,$\sigma$. We then computed the contribution of these physically related sources to the single-dish flux density. The easiest way to proceed would be to sum the flux density of all the galaxies at the same redshift and closer than a given distance. Unfortunately, this definition is problematic, since the result will depend significantly on the chosen search radius. We thus chose the following alternative method. For every galaxy at the same redshift, we computed their contribution at the center of the single-dish source by multiplying their flux density in the simulated catalog by $\textrm{exp}(-d^2 / 2\,\sigma_{\rm beam}^2)$, where $d$ is the distance between the galaxy and the center of the single-dish source and $\sigma_{\rm beam}$ is the size of the Gaussian beam. We finally divided the sum of the contribution of all these sources at the same redshift as the brightest source by the measured single-dish flux density measured in the simulated map. The results are presented in Fig.\,\ref{fig:multi}.

At 250\,$\mu$m, 80\, to 90\,\% of the flux density is emitted by the brightest galaxy. \citet{Scudder2016} found $\sim$50\,\% based on a Bayesian source-extraction method using shorter wavelength priors (black downward-facing triangles). These results could seem to contradict our analysis. However, they used a very different definition of the flux density fraction. They divided the flux density of brightest galaxy by the sum of the flux density of all the galaxies in a 1 FWHM radius. In our simulated catalog, this sum is larger than the flux density measured in the simulated map. Indeed, the numerous faint sources are responsible for a background \citep{Dole2003}, which is removed by photometric tools, and thus do not contribute to the flux densities measured in our simulated maps. In addition, the galaxies at 1\,FWHM from the \textit{Herschel} sources can contribute to another close single-dish source. Using the same method as \citet{Scudder2016}, we find a similar value of 50\,\%. However, the trend with the flux density is different. We find a rising trend, while they have a decreasing one. Their observational method is based on several important assumptions and only high-resolution far-infrared observations will allow to identify which are the most reliable. Finally, we estimated the average contribution of the other sources at the same redshift and found 5\,\%. The sum of the flux density of the brightest galaxy and other galaxies at the same redshift  remains smaller than unity. There is thus a significant contribution of galaxies at different redshifts than the brightest galaxy to SPIRE 250\,$\mu$m sources.

At 500\,$\mu$m, resolution effects are much stronger and on average only 58\,\% of the flux density is coming from the brightest galaxy. At 70\,mJy, this fraction is compatible with unity.  The $>$60\,mJy SPIRE sources are essentially local star-forming objects and lensed galaxies, which are sufficiently bright to be detected by themselves. The clustering of nearby objects is weaker than at high redshift. The contrast between a magnified source and its unlensed environment is also high. This explains why these galaxies have a smaller contamination from other galaxies when their flux density is measured with a single dish. The contribution to the measured flux density from the physically related neighbors is $\sim$10\,\% between 20\,mJy and 40\,mJy and decreases to 5\,\% for these bright sources, in agreement with our understanding.

At 1.2\,mm, we produced predictions for the NIKA2 camera \citep[e.g.,][]{Calvo2016}. The data will be less affected by resolution effects. The contribution of the brightest galaxy to the NIKA2 sources is $\sim$95\,\% at all flux densities, except the faintest ones that are close to the confusion limit. The resolution effects will be smaller with NIKA2, essentially because of the smaller beam ($\sim$12\,arcsec). The contribution from galaxies at the same redshift is $\sim$5\,\% and this fraction does not evolve significantly with the flux density. At all flux densities, the sum of the brightest galaxy and other galaxies at the same redshift is responsible for at least 97\,\% of the single-dish flux density measured in our simulated map. The contamination by low-redshift galaxies will thus be much smaller than with \textit{Herschel}, because they are observed far from their peak of emission.

\subsection{The importance of a SFR limit}

\begin{figure}
\centering
\includegraphics{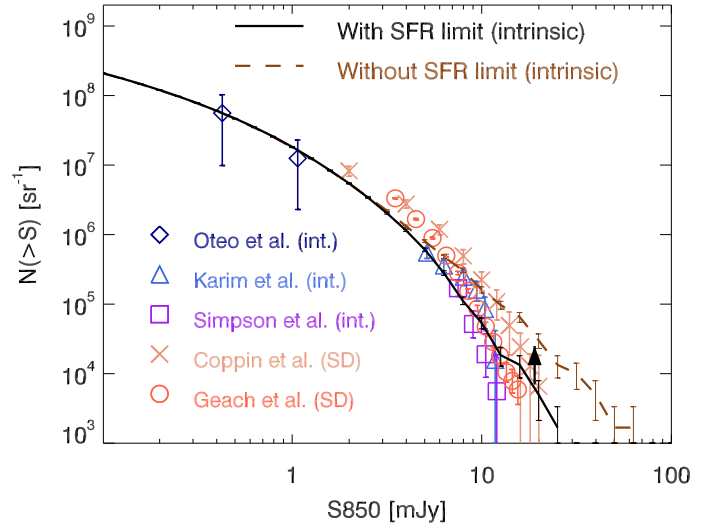}
\caption{\label{fig:sfrlim} Effect of the SFR cut on the number counts at 850\,$\mu$m. The data points are similar to those of Fig.\,\ref{fig:submmcounts}. The black solid line is our standard model, that is, with a SFR limit of 1000\,M$_\odot$/yr, and the brown dashed line is the model without SFR limit.}
\end{figure}

\label{sect:sfrlim}

In Sect.\,\ref{sect:sfrprop}, we introduced a SFR limit at 1000\,M$_\odot$/yr. The impact on number counts below 500\,$\mu$m is moderate: the model without SFR limit slightly overproduces the number of sources above 100\,mJy. On the contrary, the impact is much stronger at 850\,$\mu$m, as shown in Fig.\,\ref{fig:sfrlim}. This is not surprising because longer wavelengths are dominated by higher redshifts, where the sSFR is on average higher and more sources are thus affected by this limit. The models with and without SFR limit start to diverge at 4\,mJy. Above 10\,mJy, the model without SFR limit is 5\,$\sigma$ above the counts of \citet{Geach2017}, which should already be taken as upper limits since they are extracted from single-dish observations. The version of the model without SFR limit is thus clearly ruled out, proving a posteriori the necessity to introduce this threshold.

The SFR limit used in our simulation is an effective way to obtain number counts at the bright end in agreement with observations in the submillimeter. Other modifications could have produced similar number counts. Without SFR limit, the S$_{850}>10$\,mJy galaxies in our simulation are massive ($<$M$_\star$$>$ = 8.6$\times$10$^{10}$\,M$_\odot$) and at relatively high redshift ($<$z$>$ = 2.9). A smaller number density of massive star-forming galaxies could thus have a similar impact on the number counts. Because of the steepness of the SMF at the high-mass end, the uncertainties on the stellar mass measurements could produce an artificial excess of massive objects (an effect similar to the Eddington bias). However, it was taken into account by \citet{Davidzon2017} in their fit of the SMF. Some massive passive galaxies could also have been wrongly classified as star forming. However, the main sequence measured by stacking of star forming galaxies would also be lower. Finally, the boost of star formation in starbursts (5.3 in our simulation following \citealt{Schreiber2015}) could be lower in massive galaxies at z$>2$. A lower boost or a SFR limit causing a truncated sSFR distribution are very hard to disentangle with the current data. We thus chose the SFR-limit solution for its simplicity.

The infrared luminosity functions at z$>$2 measured with \textit{Herschel} contain objects above 10$^{13}$\,L$_\odot$ (SFR$>$1000\,M$_\odot$) even if their density drops quickly above this luminosity \citep[e.g.,][]{Gruppioni2015,Mancuso2016b}. We showed in Sect.\,\ref{sect:multi} that the SPIRE fluxes could be overestimated because of resolution effects. This could propagate to the luminosity function as discussed in Sect.\,\ref{sect:counts_zslice}. Most of the interferometric follow-up observations of these abundantly star-forming objects were performed at $\lambda > $850\,$\mu$m. Future ALMA band-9 observations (450\,$\mu$m) would thus be valuable for confirming the measurements of their obscured SFR.

\subsection{Impact of clustering on stacking analysis}

\label{sect:stacking}

Since confusion limits the detection of faint individual galaxies with single-dish instruments, a large fraction of the far-infrared and millimeter observables were measured by stacking analysis. Stacking analysis can also be biased by clustering effects. Since galaxies are clustered, there is a higher probability of finding a source in the beam of a stacked source than at a random position \citep[e.g.,][]{Marsden2009,Bethermin2010b}. Consequently the average flux density of a galaxy population measured by stacking tends to be biased toward higher values. This bias was extensively discussed in the literature and various methods were proposed to correct for this effect \citep[e.g.,][]{Marsden2009,Kurczynski2010,Bethermin2010b,Viero2013b, Heinis2013, Welikala2016}.

Our simulation is built using two observational studies based on stacking: the evolution of the main sequence measured by \citet{Schreiber2015} and the evolution of the SEDs presented in \citet{Bethermin2015a}. These results were corrected for the clustering bias using empirical approaches. Since they are key elements in the calibration of our simulation, we checked that these empirical corrections are consistent with the biases we measure in our simulation. We discuss only \textit{Herschel}/SPIRE data, since shorter wavelengths have a negligible bias ($<$10\,\%, \citealt{Bethermin2015b,Schreiber2015}). As detailed in Appendix\,\ref{app:stacking}, our values of the excess of flux density caused by clustered neighbors agree well with the estimate of \citet{Schreiber2015}: 13$\pm$1\,\% in our simulation versus $14_{-9}^{+14}$\,\% in theirs at 250\,$\mu$m, 21$\pm$1\,\% versus $22_{-14}^{+19}$\,\% at 350\,$\mu$m, and 34$\pm$1\,\% versus $39_{-23}^{+22}$\,\%. In \citet{Bethermin2015a}, we used a redshift-dependent correction estimated using two different techniques, which also agrees with our simulation as explained in the Appendix\,\ref{app:stacking}. The observables derived from a stacking analysis corrected from clustering were thus paradoxically more reliable than the statistical properties derived from catalogs of individually-detected sources.

\section{Redshift-dependent observables and consequences on the star formation history}

\label{sect:redshift}

In this section, we compare the results of our simulation with redshift-dependent observables (redshift distributions, number counts per redshift slice) and discuss the impact of these results on the determination of the obscured star formation history.

\subsection{Comparison with observed redshift distributions}

\label{sect:Nz}

\begin{figure}
\centering
\includegraphics{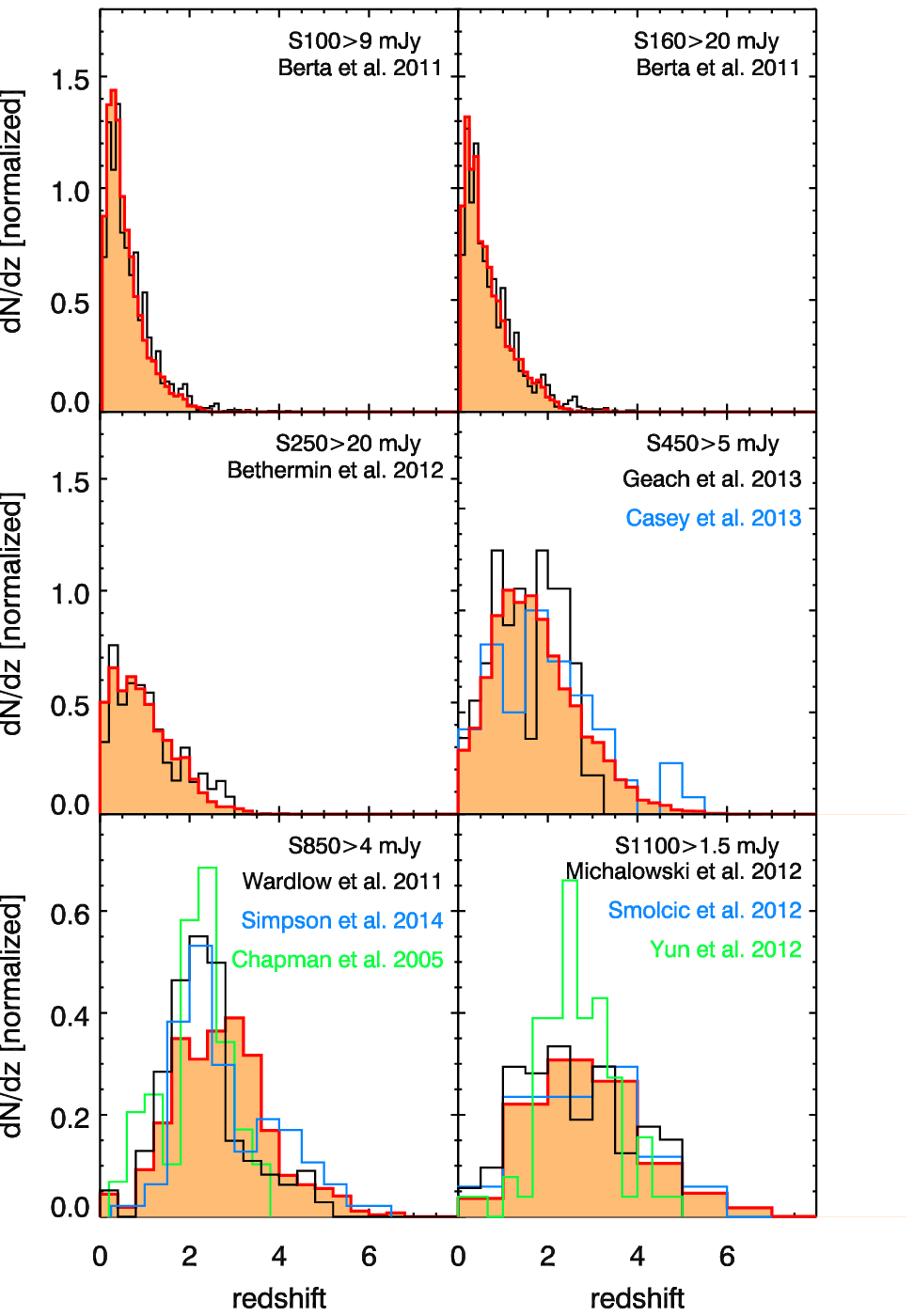}
\caption{\label{fig:Nz}Comparison between the measured redshift distributions and the predictions of our simulation. The orange histograms are the intrinsic redshift distributions from our simulation. The data points are extracted from \citet{Berta2011} at 100 and 160\,$\mu$m, \citet{Bethermin2012b} at 250\,$\mu$m, \citet{Geach2013} and \citet{Casey2013} at 450\,$\mu$m, \citet{Wardlow2011}, \citet{Simpson2014}, and \citet{Chapman2005} at 850\,$\mu$m, and \citet{Michalowski2012}, \citet{Smolcic2012}, and \citet{Yun2012} at 1.1\,mm. Figure adapted from \citet{Bethermin2015b}.}
\end{figure}

\begin{figure}
\centering
\includegraphics{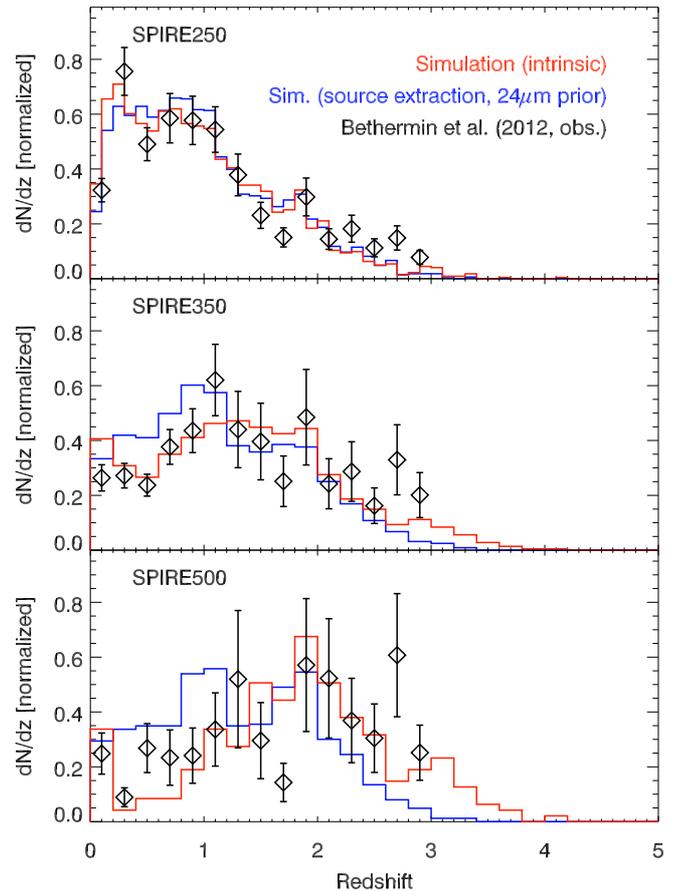}
\caption{\label{fig:Nzprior}Illustration of the impact of the 24\,$\mu$m-prior extraction on the redshift distribution at 250, 350, and 500\,$\mu$m. A flux density cut of 20\,mJy was used to select the SPIRE sources. The red line is the intrinsic redshift distribution, while the blue line is obtained after extracting the sources in our simulated \textit{Herschel}/SPIRE maps using 24\,$\mu$m positions as a prior. The histograms are normalized in order to have $\int dN/dz\,dz = 1$. We compare these model predictions with the observational redshift distribution of \citet[][black diamonds]{Bethermin2012b}. The data points were extracted in real \textit{Herschel} data using position, 24-$\mu$m flux density, and redshift as a prior. This more complex method was chosen to avoid the potential biases associated with a source extraction using only 24-$\mu$m positions as a prior (see discussion in Sect.\,\ref{sect:Nz}).}
\end{figure}

In our simulation, we implemented significant modifications compared to the \citet{Bethermin2012c} version of the model as the updated evolutions of the SEDs and of the SFR-M$_\star$ relation. We thus checked if this updated model reproduces correctly the observed redshift distributions in Fig.\,\ref{fig:Nz} (see \citealt{Bethermin2015b} for a detailed discussion about the modeling of the redshift distributions). There is an overall good agreement between the intrinsic redshift distributions in our simulation and the measured ones from 100\,$\mu$m to 1.1\,mm.

However, the measurement of the redshift distributions is a complicated task, which requires identification of the galaxy responsible for the main fraction of the far-infrared or submillimeter flux and measurement of its redshift. Various methods can be used. The procedures based on high-resolution follow-up are difficult to reproduce with our simulation. In contrast, our simulation is perfectly suited to testing the prior-based source extraction, which was used to derive \textit{Herschel} redshift distributions \citep[e.g.,][]{Berta2011,Bethermin2012b}. 

In Fig.\,\ref{fig:Nzprior}, we compared the intrinsic redshift distribution in our simulated catalog with the redshift distribution of the sources extracted in our simulated map using 24\,$\mu$m positions as a prior as described in Sect.\,\ref{sect:simobs}. At 250\,$\mu$m, the two distributions are very similar, showing this extraction technique does not bias the results. We obtained similar results at shorter wavelength with \textit{Herschel}/PACS. At 350\,$\mu$m, we found that the redshift distribution derived from the source extracted in the simulated map is slightly biased toward lower redshifts compared to the intrinsic distribution. At 500\,$\mu$m, this bias becomes stronger. As discussed in Sect.\,\ref{sect:multi}, the flux density of 500\,$\mu$m sources is emitted by several galaxies. In addition, the 24\,$\mu$m/500\,$\mu$m color varies more with redshift than colors between 24\,$\mu$m and shorter wavelengths. The brightest 24\,$\mu$m galaxy in a 500\,$\mu$m beam is thus not systematically the main contributor to the 500\,$\mu$m flux density. In conclusion, the \textit{Herschel} redshift distribution extracted using 24\,$\mu$m positional priors are thus accurate only below 250\,$\mu$m. At longer wavelength, other methods must be used.

In \citet{Bethermin2012b}, we used a prior-based extraction based on both the 24\,$\mu$m flux density and the redshift. Instead of directly selecting the brightest 24\,$\mu$m source in a 0.5\,FWHM radius as an input for FASTPHOT, we predicted the 500\,$\mu$m flux from the 24\,$\mu$m flux density and the redshift, and kept in the prior list the galaxy with the highest predicted flux density at 500\,$\mu$m in a 0.5\,FWHM radius. For the prediction, we used the average colors measured by stacking. We thus kept more high-redshift sources in the prior list. This approach agrees with the intrinsic redshift distribution in our simulation (Fig.\,\ref{fig:Nzprior}), but not with the extracted one. This highlights that more advanced prior-based source extraction techniques could be sufficient to derive accurate redshift distributions from confusion-limited maps. Our simulation will be particularly useful for validating future studies of the redshift distributions.

\begin{figure*}
\centering
\includegraphics{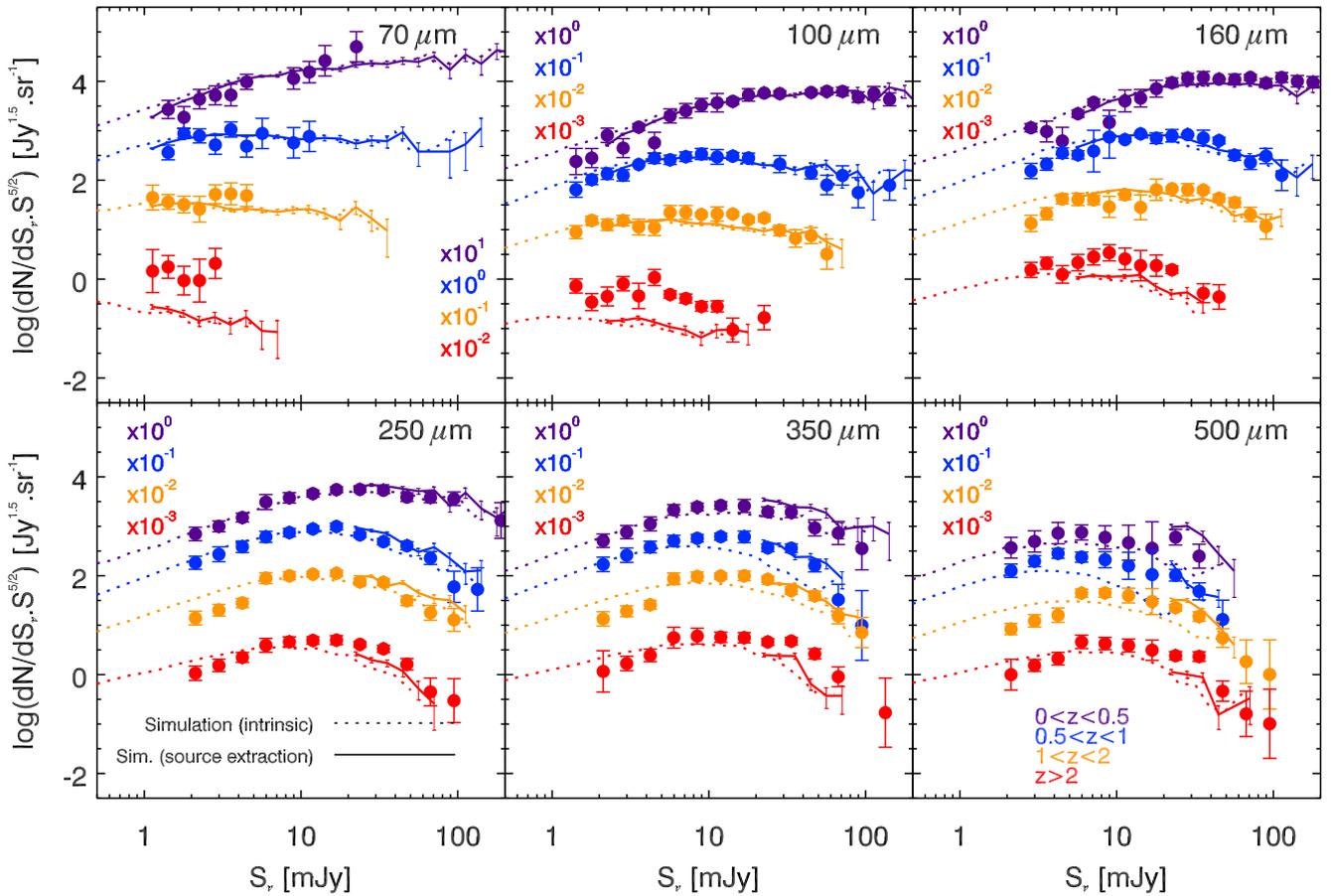}
\caption{\label{fig:counts_zslice} Number counts per redshift slice. Purple, blue, orange, and red are used for the z$<$0.5, 0.5$<$z$<$1, 1$<$z$<$2, and z$>$2 slices, respectively. The data points are from \citet{Berta2011} at 70, 100, and 160\,$\mu$m and \citet{Bethermin2012b} at 250, 350, and 500\,$\mu$m. The dotted lines are the intrinsic distributions in the simulation and the solid lines are the counts extracted using 24\,$\mu$m priors (see Sect\,\ref{sect:counts}). Figure adapted from \citet{Bethermin2012b}.}
\end{figure*}

\subsection{Number counts per redshift slice}

\label{sect:counts_zslice} 

In Fig.\,\ref{fig:counts_zslice}, we compare the results of our simulation with the measured number counts per redshift slice measured with PACS \citep{Berta2011} and SPIRE \citep{Bethermin2012b}. This observable is very close to monochromatic luminosity functions \citep{Gruppioni2013,Magnelli2013}, but is not affected by the assumptions made on the K-corrections, which are necessary to determine the luminosity functions. The full observational process used to measure the number counts per redshift slice is thus easier to simulate. There is an overall good agreement between 70 and 160\,$\mu$m at z$<$2. At z$>2$, our simulation under-predicts the source counts at 70\,$\mu$m and 100\,$\mu$m by a factor of 5 and 2.5, respectively. The number counts in our simulated catalog and after simulating the full source-extraction procedure are similar. This is thus not a problem caused by the resolution. The most likely explanation is contamination by active galactic nuclei (AGNs), since these \textit{Herschel} bands at z$>$2 correspond to $<$23 and $<$33\,$\mu$m rest-frame. Indeed, we did not implement the contribution of AGNs to the mid-infrared emission in our simulation, which focuses on the far-infrared and millimeter domain. However, at these wavelengths, 99\,\% of the sources lie at z$<$2 and the AGN contribution to the SEDs has thus a negligible impact on the global statistical properties of the galaxies.

At 250\,$\mu$m, the number counts in our simulation at z$<$2 agree well with observations and there is no significant difference between the intrinsic number counts and those extracted using 24\,$\mu$m priors. At z$>$2, the intrinsic counts under-predict the observations at the bright end, but the number counts extracted from the simulated map agree well with the data. At 350\,$\mu$m and 500\,$\mu$m, the intrinsic number counts are systematically below the observations at z$>$0.5, but the number counts extracted from the simulated maps agree better with the data. However, the number counts extracted from the simulated map tend to be lower than the observations at z$>$2 and higher at z$<$0.5. We note however, that the source extraction from the simulated map was done using only the 24\,$\mu$m position as a prior and is thus slightly biased toward low redshift, as shown in Sect.\,\ref{sect:Nz}.


\begin{figure}
\centering
\includegraphics{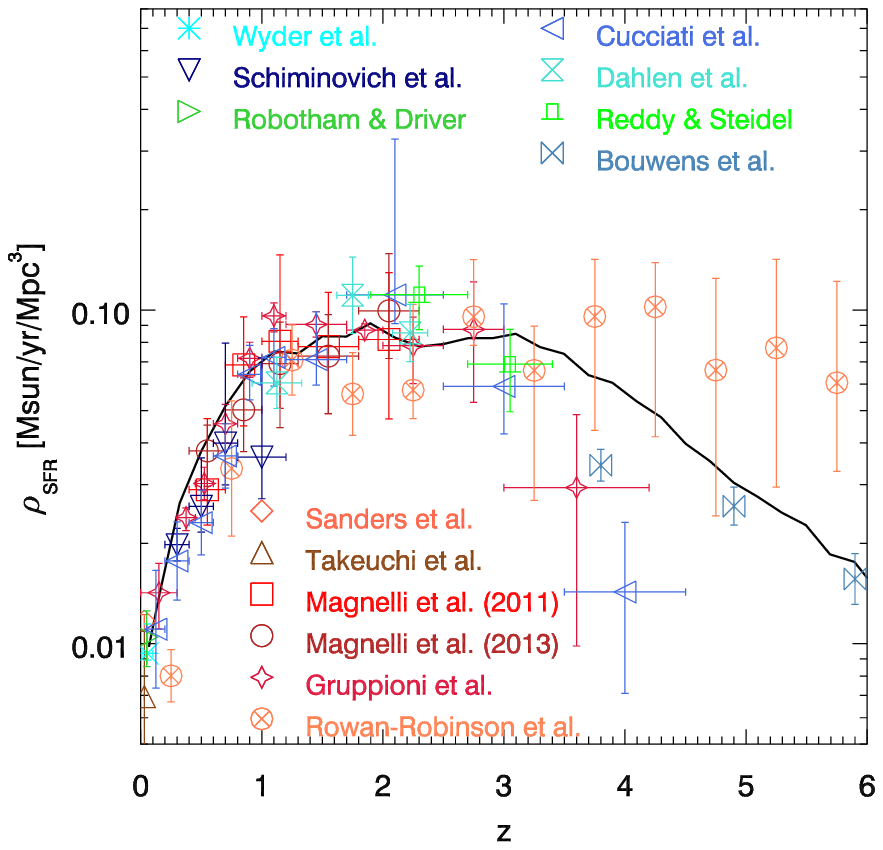}
\caption{\label{fig:sfrd}Evolution of the obscured star formation density as a function of redshift. The black line is the result of our simulation. We show for comparison the infrared measurements of \citet[diamond]{Sanders2003}, \citet[triangle]{Takeuchi2003}, \citet[squares]{Magnelli2011}, \citet[circles]{Magnelli2013}, \citet[stars]{Gruppioni2013}, and \citet[circles with cross]{Rowan-Robinson2016}. We also plot the UV estimates corrected for dust attenuation from \citet[asterisk]{Wyder2005}, \citet[downward-facing triangles]{Schiminovich2005}, \citet[right-facing triangle]{Robotham2011}, \citet[left-facing triangle]{Cucciati2012}, \citet[hourglass]{Dahlen2007}, \citet[hat]{Reddy2009}, and \citet[bowtie]{Bouwens2012b,Bouwens2012a}. All the data have been converted to \citet{Chabrier2003} IMF and \citet{Planck2015_cosmo} cosmology.}
\end{figure}

\subsection{Consequences on the obscured star formation history}

\label{sect:sfrd}

In the previous sections, we have shown that the flux densities of individually-detected sources (Sect.\,\ref{sect:counts_zslice}) are biased toward higher values because of angular resolution effects, while stacking-derived observables were already corrected from the clustering effects. Since the peak of the far-infrared emission of galaxies is around 100\,$\mu$m rest-frame, \textit{Herschel}/SPIRE data are thus essential to derive accurate obscured SFR, but unfortunately they are affected by these resolution effects. They are also limited by the confusion and only sources brighter than $\sim$20\,mJy can be extracted reliably from the maps. These bright, individually detected sources have an important role in understanding the evolution of the massive systems, but they contribute only marginally to the global star formation budget. 

At 250\,$\mu$m, the resolution has an impact only at z$>$2 (Fig.\,\ref{fig:counts_zslice}). In our simulated catalog, at z$>$2, the S$_{250} > 20$\,mJy galaxies contribute to only 2.5\,\% of the obscured star formation density. At 350 and 500\,$\mu$m, the galaxies brighter than 20\,mJy host only 2.9 and 1.4\,\% of the SFRD, respectively, at z$>$2. At those fluxes, the excess of flux density caused by the resolution effects (see Sect.\,\ref{sect:multi} and Fig.\,\ref{fig:multi}) is 21, 46, and 96\% at 250, 350, and 500\,$\mu$m, respectively. It is hard to propagate this effect to the estimate of the total infrared luminosity density and star formation density (SFRD), since it requires combining several wavelengths. However, even in the worst case scenario of using only 500\,$\mu$m as a SFR estimator, the excess of SFRD caused by sources brighter than 20\,mJy will remain below 10\,\%. This effect remains thus below the systematic uncertainties associated with the extrapolation of the contribution of the faint sources.

We checked if the SFRD in our simulation agrees or not with other estimates from the literature. In Fig\,\ref{fig:sfrd}, we compare the obscured SFRD from our simulation with the latest observations compiled by \citet{Madau2014}. Our simulation agrees well with both the IR- and UV-derived measurements up to z$\sim$3. This confirms that the impact of resolution effects are minors on the global star formation budget.

At z$>$3, our simulation is 2\,$\sigma$ higher than the measurements of \citet{Gruppioni2013} derived from \textit{Herschel} observations. However, they have only three data points at L$_{\rm IR}>10^{12.5}$\,L$_\odot$ and have to make strong assumptions about the faint-end slope of the luminosity function. In contrast, our simulation is 0.5\,$\sigma$ and 2\,$\sigma$ lower than the estimate of \citet{Rowan-Robinson2016} at z=4 and z=6, respectively. There is thus significant tension between the various estimates of the obscured SFRD at z$>3$. Our simulation agrees with the measured redshift distributions and deep millimeter counts and is thus compatible with the current non-extrapolated data. These differences between studies highlight how uncertain the obscured star formation history at z$>$3 remains. Concerning the observation derived from dust-corrected UV, our simulation agrees with \citet{Bouwens2012b,Bouwens2012a} measurement at z$>$5, but is 50\,\% higher at z$\sim$4. This suggests that a fraction of the star formation might have been missed at this redshift by optical surveys. Future wide and deep millimeter surveys with NIKA2 at IRAM and with the large millimeter telescope (LMT) will be essential to confirm or not this result.

\section{One- and two-point map statistics}

\label{sect:pdpk}

In addition to the statistical properties of the sources, we checked the agreement of our model with map statistics. This is particularly important for SPIRE data, which has a limited resolution.

\begin{figure}
\centering
\includegraphics{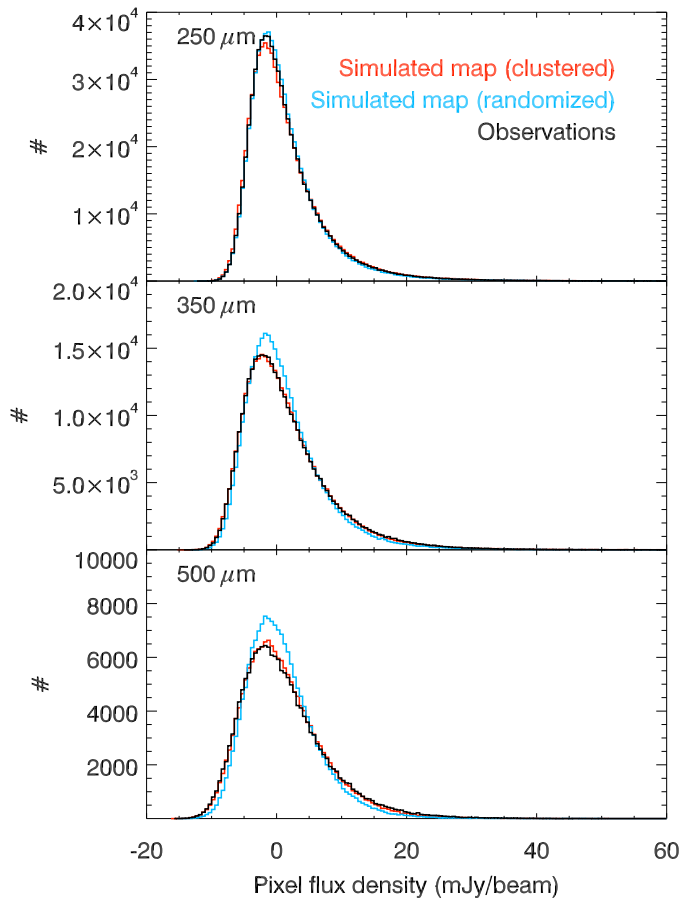}
\caption{\label{fig:pd} Pixel histograms of the \textit{Herschel}/SPIRE maps in COSMOS (black) and comparison with our simulation using the same instrumental noise map. The red histograms are the result of our simulation and the blue ones are the histogram obtained after randomizing the position of the sources to illustrate the impact of clustering. 
}
\end{figure}

\begin{figure*}
\centering
\includegraphics[width=18cm]{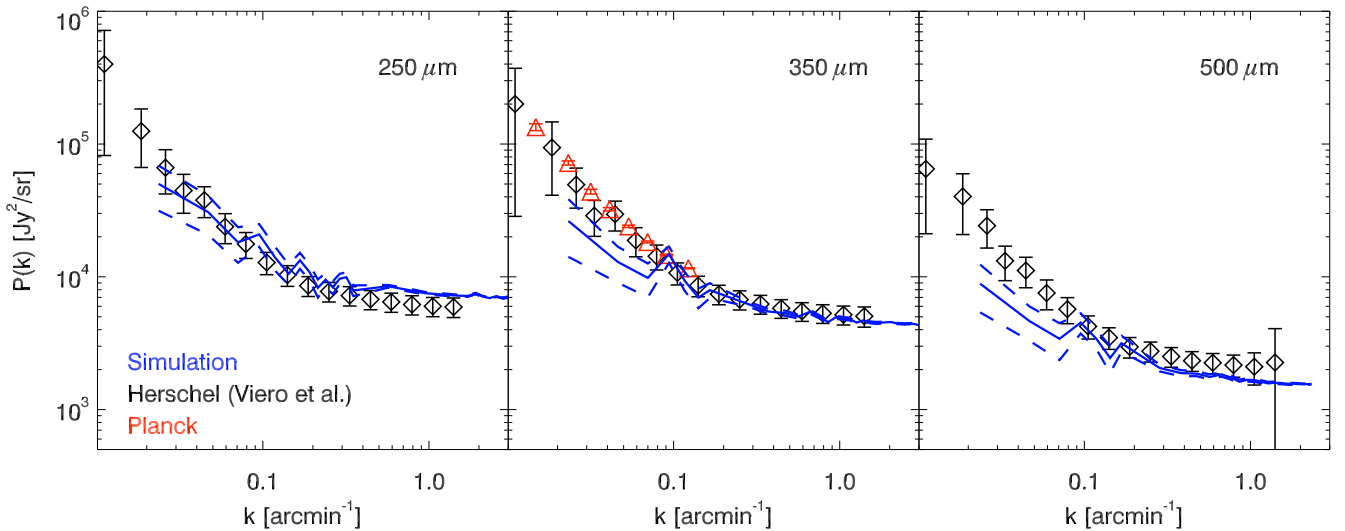}
\caption{\label{fig:pk} Power spectrum of cosmic infrared background anisotropies at 250, 350, and 500\,$\mu$m. The blue solid lines are CIB anisotropies measured in our simulation and 1\,$\sigma$ confidence regions are represented by the dashed lines. The black diamonds and the red triangles are the measurements of \citet{Viero2013} and \citet{Planck_CIB2013}, respectively.}
\end{figure*}

\subsection{Pixel histograms: P(D)}

The distribution of the surface brightness in the pixels (P(D)) of a map is directly connected to the number counts of the objects in this map \citep{Scheuer1957,Condon1974}. This method was used to measure the faint source counts with, for example, Bolocam \citep{Maloney2005}, LABOCA \citep{Weiss2009}, BLAST \citep{Patanchon2009}, and \textit{Herschel} \citep{Glenn2010}. As discussed in \citet{Takeuchi2004} and \citet{Patanchon2009}, clustering could impact P(D) analysis. However, when these analyses were performed, simulations did not include clustering that we know is critical. We can now investigate this effect using our new simulations.

In Fig.\,\ref{fig:pd}, we compare the pixel histograms of the simulated maps and of real \textit{Herschel} maps. We used the COSMOS maps\footnote{\url{http://hedam.lam.fr/HerMES/}} from the HerMES survey \citep{Oliver2012}, which match the size of our simulation. We used the real noise maps released by the HerMES team to generate a similar Gaussian instrument noise in our simulated maps. In order to evaluate the impact of clustering, we produce another simulated map without clustering by randomly reshuffling the positions of the galaxies. 

At 250\,$\mu$m, the clustering has an impact of less than 5\,\% and both clustered (red) and unclustered (blue) simulated maps agree at 5\,\% with the observed histogram. This is a very good agreement considering the 4\,\% calibration uncertainty of SPIRE \citep{Bendo2013}. At 350 and 500\,$\mu$m, the effect of clustering is much larger because of the larger beam and can reach 15\,\%. The clustered maps agree well with the observed one, but the randomized maps have a large excess at the peak. This shows that clustering has a non-negligible effect on the P(D) analysis and must be taken into account. This could explain why the P(D) analysis of \citet{Glenn2010} agrees with individual source counts even if they are biased high compared to the intrinsic counts (see Fig.\,\ref{fig:herschel_counts}).


\subsection{Anisotropies of the cosmic infrared background: P(k)}

Measuring the clustering of individually detected population in confusion-limited data is difficult. The sample sizes remain limited to obtain good statistics \citep[e.g.,][]{Bethermin2014}. The contamination of the fluxes by the neighbors tends to introduce artificial correlation between redshift slices and to bias the measurements \citep{Cowley2016}. The power spectrum of the CIB anisotropies, which is not affected by this problem, is currently the best way to constrain how the star formation is distributed in dark-matter halos \citep[e.g.,][]{Lagache2007,Bethermin2013,Planck_CIB2013,Viero2013}. CIB anisotropies are a powerful observation to test that the model
simultaneously reproduces the infrared emission and the spatial distribution of galaxies.

In Fig.\,\ref{fig:pk}, we compare the power spectrum measured with \textit{Herschel} \citep{Viero2013b} (black diamonds) and in our simulation (blue solid lines). In order to reduce the Poisson noise, the brightest sources are usually masked or subtracted from the maps. We chose to use a S$_{\nu, \rm cut}=50$\,mJy flux density cut, which is the deepest cut used by \citet{Viero2013b}. We also included the \textit{Planck} data at 857\,GHz (350\,$\mu$m, red triangles). We shifted these data by the difference of the Poisson noise in our simulations between a flux density cut of 50\,mJy and 710\,mJy (used by \textit{Planck}). Finally, as shown by \citet{Bertincourt2016}, the \textit{Herschel}/SPIRE 500\,$\mu$m absolute flux calibration is 4.7\,\% too high compared to \textit{Planck}. We thus corrected \citet{Viero2013b} data points accordingly. To
accurately measure the power spectrum, we generate a map without convolving it by the PSF and without including the sources above the flux cut. To be fully consistent with the observational process, we produced a map using the SPIRE spectral response to extended emission to measure the power spectrum, but we used the flux densities from the point-source spectral response to select the sources to put in the map (see Lagache et al. in prep. for a detailed discussion). We measured the power spectrum from these simulated maps using the POKER software \citep{Ponthieu2011,Planck_CIB2011,Planck_CIB2013}. This software accounts for non-periodic boundary conditions of the map that otherwise bias large-scale measurements. The error bars are estimated via Monte-Carlo simulations of the estimated power spectrum.

At small scale (k$>$0.3\,arcmin$^{-1}$), the power-spectrum is dominated by the shot noise from galaxies. These Poisson fluctuations of the number of galaxies in a patch of sky produce a plateau in the power spectrum, which can be derived directly from the number counts \citep{Lagache2000}:
\begin{equation}
\label{eq:poisson}
\sigma_{\rm Poisson}^2 = \int_0^{S_{\nu, \rm cut}} S_\nu^2 \, \frac{d^2 N}{dS_\nu d\Omega} \, dS_\nu,
\end{equation}
where $S_\nu$ is the flux density and $\frac{d^2 N}{dS_\nu d\Omega}$ are the differential number counts. At 350 and 500\,$\mu$m, our simulation agrees at 1\,$\sigma$ with the measurements of \citet{Viero2013b}. At 250\,$\mu$m, our simulation is systematically 1.5\,$\sigma$ above the measurements. Since the measurements are dominated by systematic effects (e.g., deconvolution of the beam), their error bars are strongly correlated. This offset is thus not statistically significant. Overall, the Poisson level in our simulation and in real data agrees.

\citet{Mak2017} found a discrepancy between the Poisson level measured in \textit{Herschel} and \textit{Planck} data and that derived from the measured \textit{Herschel} number counts using Eq.\,\ref{eq:poisson}. The Poisson level derived directly from the number counts are higher than the measurements. This problem can be solved naturally considering the discrepancy between the measured and the intrinsic number counts that we identified in Sect.\,\ref{sect:herschel_counts}. Indeed, the Poisson level depends on the number counts. Since the observed number counts are overestimated at 350 and 500\,$\mu$m because of resolution effects, the Poisson levels derived from them are thus overestimated.

At 250\,$\mu$m, our simulation agrees at better than 1\,$\sigma$ with the data at large scale (k$<$0.3\,arcmin$^{-1}$). In this regime, the power spectrum is dominated by the large-scale clustering of galaxies. At 350 and 500\,$\mu$m, at k$<$0.1 arcmin$^{-1}$, our simulation underestimates the power spectrum by 1.5\,$\sigma$. Future larger simulations will allow us to determine if this deficit is real or just a statistical fluctuation.

\section{The nature of \textit{Herschel} red sources}

\label{sect:red}

\citet{Dowell2014} and \citet{Asboth2016} found a large population of red \textit{Herschel} sources in the HerMES survey \citep{Oliver2012}. They claimed that the number of sources they found is one order of magnitude higher than predicted by the models. If confirmed, these results would suggest that the models strongly underpredict the number of bright z$>$4 dusty star forming objects.  \citet{Ivison2016} also found a large number of red high-redshift candidates in the H-ATLAS survey using a slightly different selection.

In this section, we verify that our simulation accurately reproduces the statistics of red sources. First, we check the statistics of red sources in our 2\,deg$^2$ simulation, which is a small area but includes clustering (Sect.\,\ref{sect:red_small}). We then investigate the statistics of red sources in a large simulated catalog, without clustering (Sect.\,\ref{sect:red_large}). 
The criteria of \citet{Ivison2016} are hard to reproduce, since they involve some visual inspection. We thus focus our analysis on the results of \citet{Asboth2016}, who used the following criteria: $S_{250}<S_{350}<S_{500}$, $S_{500} > 52$\,mJy, and D = 0.92 M$_{500}$ - 0.392 M$_{250}$ > 34\,mJy, where M$_{500}$ and M$_{250}$ are the values of the maps at the position of a source at 250 and 500\,$\mu$m, respectively, after matching all the maps at the resolution of 500\,$\mu$m data.

\subsection{Simulation in map space in 2\,deg$^2$}

\label{sect:red_small}

There is no galaxy in our 2\,deg$^2$ simulated catalog that follows the \citet{Asboth2016} criteria. However, as we will show, some of these sources could be explained by noise fluctuations and resolution effects.

\citet{Asboth2016} homogenized the beams to the size of the 500\,$\mu$m one. For simplicity, we directly generated three SPIRE maps using a Gaussian beam with a FWHM of 36.3\,arcsec. We then added instrumental noise using the same values as in \citet{Asboth2016}. The D map is generated from the 250\,$\mu$m and 500\,$\mu$m maps. The noise in our simulated D map is very close to the observations: 8.8 versus 8.5\,mJy. This shows that our approximation on the beam is sufficiently good to perform our analysis of red sources.

We extracted the peaks higher than 34\,mJy in the D map and measured the photometry on the maps at the native SPIRE resolution using the flux in the central pixel of the source after subtracting the mean of the map. The number of detected red sources varies depending on the realization of the noise. We thus used 1000 realizations to estimate the mean number of detected sources. This estimate does not take into account the cosmic variance. We found 1.7$_{-0.9}^{+1.9}$ in our 2 deg$^2$ field, which corresponds to 229$_{-121}^{+258}$ sources in 274\,deg$^2$. This agrees at 1\,$\sigma$ with the 477 detections reported by \citet{Asboth2016}. Even if the statistics are very limited, our results indicate that there might be no real tension between models and observations of red sources.

\subsection{Red sources in a 274\,deg$^2$ catalog}

\label{sect:red_large}

Our simulation covers only 2\,deg$^2$ and thus contains small statistics compared to \citet{Asboth2016} who used 274\,deg$^2$. We thus generated another simulated catalog based on the same prescriptions and covering the same area of 274\,deg$^2$ as \citet{Asboth2016}. Since we do not have a sufficiently large dark-matter simulation, we have to ignore the clustering and draw directly the sources from the stellar mass function. With this simplified method, we could potentially underestimate the number of red sources, since we neglect the boosting of the flux of massive high-redshift dusty galaxies by their neighbors. This effect can be potentially important because of the strong clustering of the most star-forming galaxies at z$>2$ \citep[e.g.,][]{Farrah2006,Bethermin2014}. We generated only sources with M$_\star > 10^{10}$\,M\,$\odot$ and z$>1$ to save memory, because no source with a lower mass can be sufficiently bright and no source below z=1 can be sufficiently red to pass the "red source" selection. We do not produce maps so we compute D from the flux densities (0.92 S$_{500}$ - 0.392 S$_{250}$). 

\begin{figure}
\centering
\includegraphics{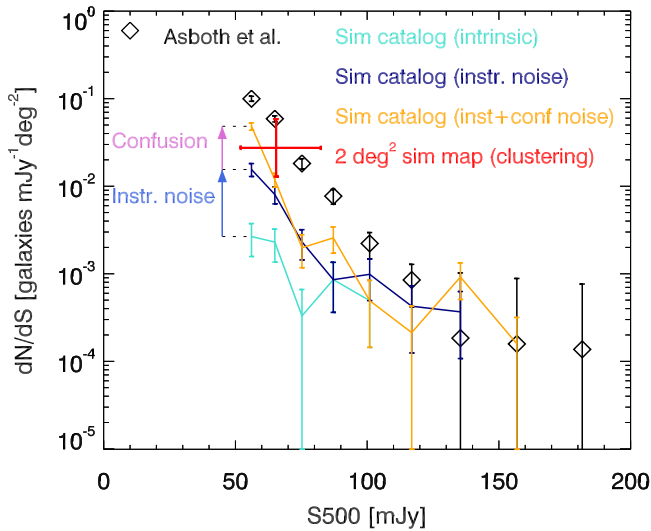}
\caption{\label{fig:red_counts} Differential number counts of red \textit{Herschel} sources. The black diamonds are the measurements from \citet{Asboth2016}. The red point is the result of an end-to-end analysis of our 2\,deg$^2$ simulated maps including clustering and instrumental noise. The turquoise, dark blue, and gold lines are the number counts in a simulated catalog of 274 \,deg$^2$ without noise, with instrumental noise only, and with both instrumental and unclustered confusion noise, respectively. Blue and purple arrows illustrate the impact of instrumental and confusion noise, respectively.}
\end{figure}

We have only 18 objects following the criteria of \citet{Asboth2016} in our simulated catalog compared to the 477\,objects in \textit{Herschel} data. All these 18 sources are strongly lensed ($\mu > 2$). Our simulation contains 439\,256 non-lensed sources with $S_{250}<S_{350}<S_{500}$, but none of them are sufficiently bright to satisfy $S_{500} > 52$\,mJy. Number counts of red sources at 500\,$\mu$m are shown in Fig.\,\ref{fig:red_counts}. While the counts from our simulation above 100\,mJy are close to the data, they are well below for fainter flux densities. The problem is thus not coming from the lensing, but from a lack of intrinsically bright sources. If we remove the SFR limit in our simulation, we find 205 non-lensed sources matching the Asboth et al. criterion (676 if we add noise to the simulated catalog, see next paragraph). However, this SFR cut is necessary to be consistent with the 850\,$\mu$m number counts (see Sect.\,\ref{sect:sfrlim}). Another explanation must thus be found to understand this discrepancy.

The results described in the previous paragraph ignore the effect of noise on the number counts of red sources. We thus simulate the effect of both confusion and instrumental noise by adding a random Gaussian noise to the flux densities of the simulated sources. We used the noise values provided in Table 1 of \citet{Asboth2016}. This method produces a higher noise on D than in the real D map, since the confusion noise from 250 and 500\,$\mu$m tend to partially cancel each other out in the real maps. Because of the very steep color distribution and number counts, the noise strongly
 increases the number of red sources (see Fig.\,\ref{fig:red_counts}). We found 168 sources matching the Asboth et al.\ criteria (74 with only instrumental noise), of which only 29\,\% are strongly lensed (60\% with only instrumental noise). The noise has thus an important role in producing red sources without strong lensing. Concerning the weakly-lensed sources, the weakly-lensed red sources are on average 6\,\% more magnified than the average magnification at z$>$2. Red-source selections are thus biased toward higher magnifications. Weak lensing acts as an additional noise on the flux density and sources on a positive fluctuation of the magnification tend to pass the 500$\mu$m flux density threshold more often. This is similar to the Eddington bias. Without weak lensing, the number of red sources decreases to 132. 

In our simulation, the noise thus strongly increases the number density of detected red sources. These results could seem to be in contradiction with \citet{Asboth2016}, who found that the number of injected and recovered red sources was similar with the number of recovered ones using an end-to-end simulation. However, we identified one potentially incorrect assumption in their simulation. Their simulation used the \citet{Bethermin2012c} model with the intrinsically red source removed and a power-law distribution of red sources with a fixed color based on the median observed one. The flux distribution of the red sources in their simulation is based on the number of detected red sources directly extracted in the real map, which is one order of magnitude higher than what is intrinsically in the \citet{Bethermin2012c} model. The relative contribution to the extracted number counts of intrinsically non-red sources, which matches the red criteria because of the noise and resolution effects, might thus be significantly underestimated because of the very high number of intrinsically red sources in their simulation. This highlights the difficulty to correct the biases in statistical measurements of red sources using only inputs from observations.

\begin{figure*}
\centering
\includegraphics{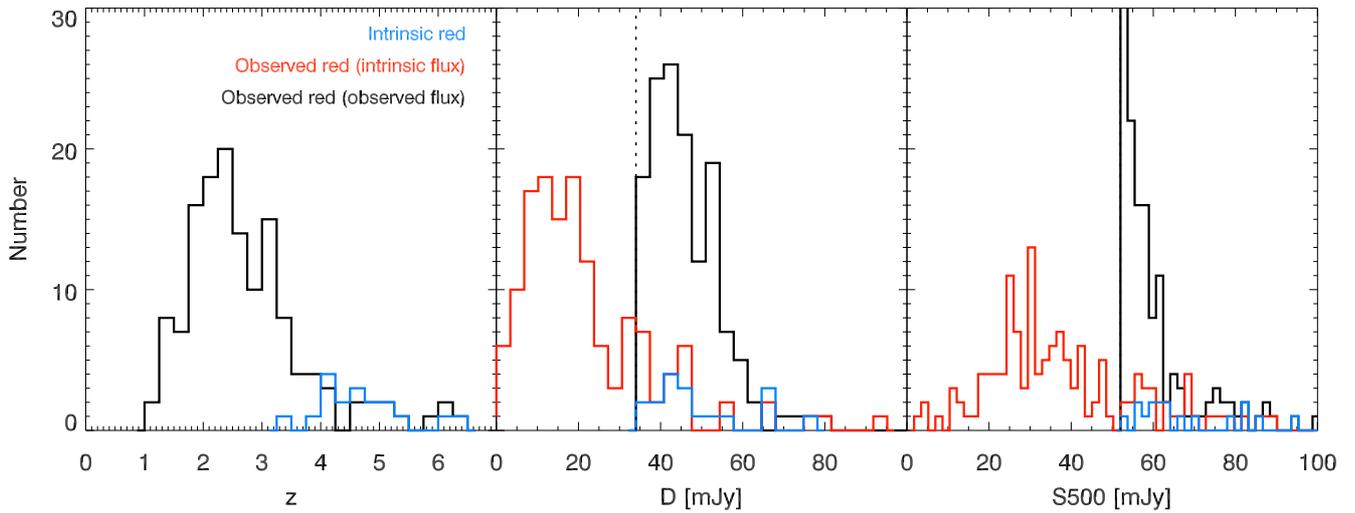}
\caption{\label{fig:red_histo} \textit{Left panel:} redshift distribution of the red sources. The blue and black histograms are the distributions of the sources matching the red criteria before and after including noise (both instrument and confusion noises), respectively. \textit{Middle panel:} distribution of the red excess  (D = 0.92 M$_{500}$ - 0.392 M$_{250}$). \textit{Right panel:} distribution of the 500\,$\mu$m flux density. The blue and black histograms show the intrinsic (without including noise) 500\,$\mu$m flux densities (middle panel) and D values (right panel) of the intrinsically red (red criteria applied before adding noise) and observed red (red criteria applied after adding noise) sources, respectively. The red histogram shows the distributions of the observed, that is, after including noise, 500\,$\mu$m flux densities and D values of observed red sources. The difference between the black and red histogram illustrates that sources observed above the S$_{500}$ and D thresholds of \citet[][vertical black dotted line]{Asboth2016} usually have lower intrinsic values.}
\end{figure*}

In Fig.\,\ref{fig:red_histo}, we illustrate the impact of the noise on the selection of red sources. If we apply the \citet{Asboth2016} criteria to the intrinsic fluxes of the sources, red sources are selected at z$>$3. If we select red sources with the same criteria after adding noise, the number of detections at z$<$4 increases dramatically, while the number of z$>$4 detections remains almost constant (see left panel). The noise has thus a strong impact on the redshift distribution of red sources. The red sources selected in the noisy catalog usually have a measured D value and a 500\,$\mu$m flux density just above the limit (vertical black dotted line in middle and right panels of Fig.\,\ref{fig:red_histo}). However, their intrinsic D value and 500\,$\mu$m flux density are in general below the detection limit ($\langle$D$\rangle = 23$\,mJy and $\langle$S$_{500} \rangle = 42$\,mJy). The intrinsic non-red sources selected because of the noise are thus not purely spurious objects. They are mostly strongly star-forming objects with a D value and an intrinsic 500\,$\mu$m flux density slightly below the cut. The first red sources that have been followed up were selected in deep fields and were usually the reddest sources in the sample \citep[e.g.,][]{Riechers2013, Dowell2014}. These objects were confirmed spectroscopically to be at high redshift. However, their selection was less affected by the noise because of their particularly red observed color and the lowest instrument noise in such fields.

\subsection{The challenge of using red sources to constrain models}

Overall, the combination of the noise and the weak lensing can
dramatically boost the number density of detected red sources compared to their intrinsic number density. Our 2\,deg$^2$ end-to-end simulation including clustering is compatible with the observations, but the statistics are limited. We also built a larger simulation based on a catalog of galaxies with random positions and a random Gaussian noise. However, the number of red sources remains lower by a factor of 2.8 in this simplified simulation. This could be explained by several effects. As shown in the pixel histogram of \textit{Herschel} maps in Fig.\,\ref{fig:pd}, the combination of instrument and confusion noise has an asymmetrical distribution with a large positive tail, which could be responsible for more bright 500\,$\mu$m outliers than in the Gaussian case. In addition, the faint foreground sources are clustered and a local underdensity of faint blue foreground galaxies in a beam could create an artificial red color. Finally, the model is sensitive to the value of the SFR cut and better observational constraints should allow us in the future to determine its value or favor one of the alternative scenarios described in Sect.\,\ref{sect:sfrlim}.
 
Comparing observations of red sources and models remains complicated, because it would require using the exact same algorithm on simulated maps with clustering of $\sim$100\,deg$^2$. For wide shallow fields as the one used by \citep{Asboth2016}, the noise also plays a crucial role. Applying a new method of deblending to the deeper 55\,deg$^2$ HeVICS field, Donevski et al. (to be submitted) found that number counts of red sources are an order of magnitude lower than \citep{Asboth2016}. They simulated the effect of instrument noise and confusion on our simulated catalogs and found an excellent agreement with their new observations. They also found that the noise has only a mild impact on the number counts of red sources in this deeper field. This highlights the need to perform end-to-end simulations with the same extraction algorithm and realistic noise properties to compare observations and model. However, these simulations remain extremely difficult to perform on large areas because they require both a large-volume dark-matter simulation and a sufficiently-high-mass resolution to have the faint galaxies responsible for confusion.
 
In addition to these difficulties, the lensing was included in a non-consistent way in our simulation, since it was drawn randomly in a distribution, which does not vary with the position of the source (see Sect.\,\ref{sect:limitations}). Non-trivial biases can occur in color selections for lensed sources. For instance, the sources clustered with the lens can change the color of the source, since they are at lower redshift and thus bluer than the background source \citep{Welikala2016}. In addition, we used only a simplified Gaussian weak lensing. The number of sources with a magnification between 1.5 and 2 is thus underestimated compared to the numerical simulation of \citet{Hilbert2007}. For instance, the extreme starbursts reported by \citet{Riechers2013} were later proved to be lensed \citep{Cooray2014} by a small factor.
 
Interpreting the statistics of red sources with models is thus a very challenging task, because of the complexity of the various artifacts affecting the selection of these objects. Direct millimeter selections as SPT \citep{Vieira2013} provides more straightforward constraints for models. However, we should mention that, despite the difficulty in interpreting their number counts, red-source selections in \textit{Herschel} fields are a powerful tool to build large samples to study the physics of high-redshift, dusty star forming galaxies. 


\section{Conclusion}

We presented a new simulation of the far-infrared and (sub)millimeter sky called SIDES. This simulation is based on an updated version of the \citet{Bethermin2012c} phenomenological galaxy evolution model using the latest observational constraints on the stellar mass function, the main sequence of star forming galaxies, and the evolution of the SEDs. To obtain realistic clustering, we used an abundance matching procedure to populate the dark-matter halos of a light cone constructed from the Bolshoi-Planck simulation \citep{Rodriguez-Puebla2016} with the galaxies produced by our model. The intrinsic galaxy number counts in this new simulation are significantly lower than the measurements  from single-dish instruments, while they agree with interferometric data. To understand this tension between our simulation and the observations, we simulated the full source extraction process and showed that the number counts extracted from our simulated maps agree with the observed ones. When we take into account the observational effects, our simulation is able to simultaneously reproduce the single-dish and interferometric number counts from the far-infrared to the millimeter domain together with redshift distributions, CIB anisotropies, and the pixel histograms of SPIRE maps.\\

Our simulation also allowed us to evaluate the impact of clustering and angular resolution on some statistical properties derived from far-infrared and (sub)millimeter surveys. We identified the following effects:
\begin{itemize}
\item The flux density of \textit{Herschel} sources is affected by resolution effects. The brightest galaxy in the \textit{Herschel} beam is responsible for $\sim$85\,\% of the flux density at 250\,$\mu$m, but only $\sim$60\,\% at 500\,$\mu$m. Other galaxies contributing to the \textit{Herschel} flux density are both galaxies at the same redshift as the brightest one and randomly aligned galaxies. Our simulation predicts that the fraction of the flux density coming from the brightest galaxy will rise to 95\,\% in future millimeter single-dish surveys performed by 30 meter-class telescopes (e.g., NIKA2 at IRAM).
\item Measurements using stacking are also biased by the clustering. However, this bias is already known. The corrections made by the observational studies are compatible with the corrections derived from our simulation. Paradoxically, the stacking studies, which took into account the clustering, are more accurate than the observations of individually detected sources.
\item The redshift distributions of \textit{Herschel} sources extracted using 24\,$\mu$m positional priors tend to be biased toward lower redshifts at 350 and 500\,$\mu$m, but are reliable at shorter wavelengths.
\item Even if the flux density of the brightest \textit{Herschel} sources tends to be overestimated, the impact of these sources on the global star formation history is small. Our simulation is compatible with the UV- and IR-derived measurements of the SFRD up to z$<$3. At z$\sim$3.5, our simulation predicts a higher SFRD than the measurements of \citet[IR]{Gruppioni2013} and \citet[UV]{Bouwens2012a,Bouwens2012b}. At higher redshift, our results are compatible with the constraints from the UV.
\item The clustering has a significant impact on the pixel histogram of \textit{Herschel} maps used to perform P(D) analysis. This explains why the number counts derived by \citet{Glenn2010} using this statistical technique are compatible with the measurements derived using standard source-extraction methods, but not with the intrinsic number counts.
\item The resolution effects allow us to solve the tension between the measured level of the Poisson fluctuation of the CIB power spectrum and what is expected from the observed number counts. Indeed, the number counts are biased high and the shot-noise derived from the number counts is thus overestimated.
\item Recently, \citet{Asboth2016} identified a population of red \textit{Herschel} sources. Their number density is one order of magnitude higher than in models, including our new simulation. However, after taking into account the noise and the resolution effects, our 2\,deg$^2$ simulation produces the correct number of objects.\\
\end{itemize}

These results highlight the difficulty to interpret the long-wavelength single-dish observations. Correcting the observations for all the observational effects is a complex task. The corrections usually assume an underlying model. In this paper, we started from our model and reproduced the full observational process. This approach is more direct and allowed us to test the validity of our model without having to rely on complex corrections of the data. This approach is probably the best way to deal with the complexity and the precision of the modern data sets.\\

Our simulation (SIDES), released publicly at \url{http://cesam.lam.fr/sides}, has many potential applications:
\begin{itemize}
\item It can be used to prepare future surveys and in particular to predict the number of detected galaxies and their properties (redshift, SFR, stellar mass). The realistic clustering included in our simulation can also be used to accurately estimate the confusion limit.
\item It is a powerful tool to test source-extraction techniques and characterize their biases. We can also use it to test and optimize methods to identify the counterparts at shorter wavelengths of single-dish sources. Finally, it can be used to validate stacking softwares and determine the most efficient and less biased ones.
\item The various biases affecting the extraction of single-dish sources can have strong impacts on clustering measurements and bias the estimates of the host halo mass \citep[e.g.,][]{Cowley2016}. Since it accurately reproduces a large set of observables, our simulation is well suited to characterize and correct for these effects.
\item Finally, in a future paper, we will include the emission of far-infrared and (sub)millimeter lines in our simulation and perform predictions for spectroscopic surveys and for (sub)millimeter intensity mapping experiments.
\end{itemize}

\begin{acknowledgements}
We thank the anonymous referee for his/her very useful and constructive comments. We thank Y.-Y. Mao for providing the abundance matching code and for assistance. We thank Eric Jullo, St\'ephane Arnouts, Olivier Ilbert, and Corentin Schreiber for their explanations. The Bolshoi-Planck simulation was performed by Anatoly Klypin within the Bolshoi project of the University of California High-Performance AstroComputing Center (UC-HiPACC) and was run on the Pleiades supercomputer at the NASA Ames Research Center. We acknowledge financial support from "Programme National de Cosmologie and Galaxies" (PNCG) funded by CNRS/INSU-IN2P3-INP, CEA and CNES, France. This work has been partially funded by the ANR under the contract ANR-15-CE31-0017. This work has been carried out thanks to the support of the OCEVU Labex (ANR-11-LABX-0060) and the A*MIDEX project (ANR-11-IDEX-0001-02) funded by the "Investissements d'Avenir" French government program managed by the ANR. HW\ acknowledges the support by the US\ National Science Foundation (NSF) grant AST1313037. Part of the research described in this paper was carried out at the Jet Propulsion Laboratory, California Institute of Technology, under a contract with the National Aeronautics and Space Administration.

\end{acknowledgements}

\bibliographystyle{aa}

\bibliography{biblio}

\begin{appendix}

\begin{figure*}
\centering
\begin{tabular}{cc}
\includegraphics{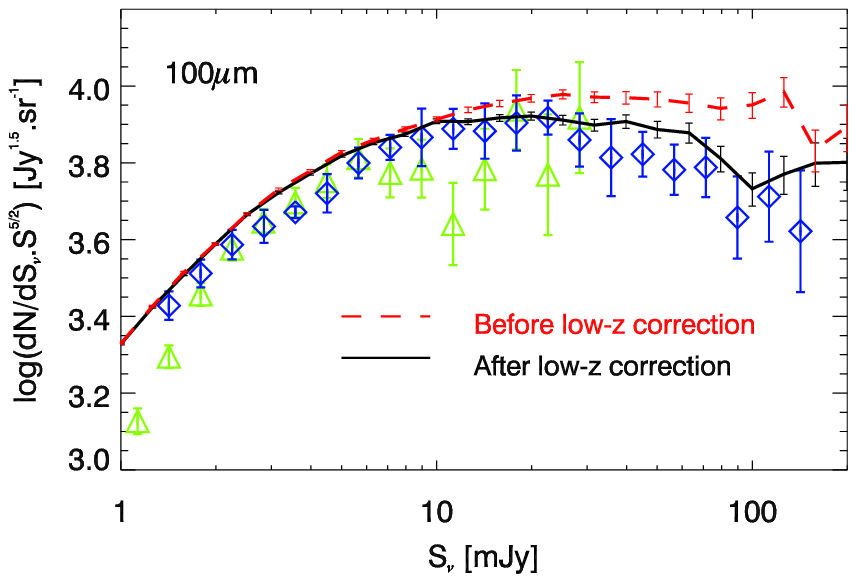} & \includegraphics{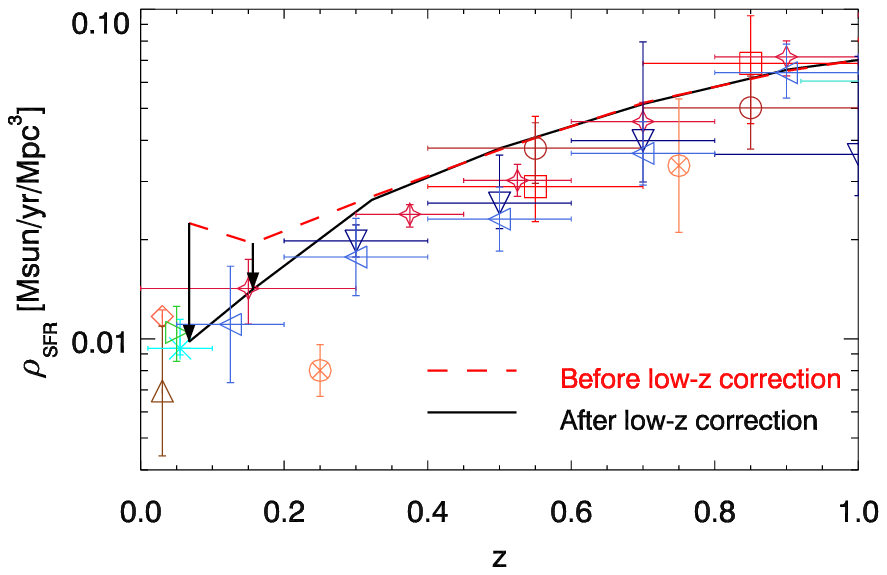} \\
\end{tabular}
\caption{\label{fig:lowz} \textbf{Left:} Differential number counts at 100\,$\mu$m. We used the same data points as in Fig.\,\ref{fig:herschel_counts}. The red dashed line is before the correction at low redshift of the position of the main sequence (see Sect.\,\ref{sect:lowzcorr}) and the black solid line is the final version of our model. Since this effect is only significant at bright flux density, where the number of sources is small, we used an unclustered 10\,deg$^2$ to reduce the statistical uncertainties in our model predictions. \textbf{Right:} Star formation density as a function of redshift. The data are similar to Fig.\,\ref{fig:sfrd}, but we added the prediction of our model before the low-redshift correction of the main sequence (red dashed line).}
\end{figure*}

\section{Homogenization of cosmology}

\label{sect:homogenization}

Our simulation is based on observational quantities (e.g., stellar masses, SFR). They were derived assuming a cosmology that was different from the one used in the dark-matter simulation. To be consistent, we convert the observational constraints to the \textit{Planck} cosmology. At fixed SED, the SFR and stellar masses are proportional to the intrinsic luminosity of an object. In the \textit{Planck} cosmology, the luminosity distance is $\sim$3\% (small redshift dependance) larger than in the 773 cosmology used in most of the observational papers ($h = 0.7$, $\Omega_\Lambda  = 0.7$, $\Omega_M  = 0.3$). The intrinsic luminosity of the objects, and consequently the stellar masses and SFRs, are thus slightly higher than in the observational papers in 773 cosmology. We thus applied a (D$_{\rm L, \textit{Planck}}$/D$_{\rm L, \textit{773}}$)$^2$ correction to the observed stellar masses and SFR. Similarly, the volume corresponding to a redshift slice and estimated with 773 cosmology is smaller than with \textit{Planck} cosmology. The number density of observations estimated in 773 cosmology is thus overestimated. We thus apply $(dV_{\rm comoving, \, 773}/dz) / (dV_{\rm comoving,\, \it Planck}/dz)$ corrections to the characteristic densities $\Phi$. The correction is computed at redshift corresponding to the center of the redshift bins used to derive the mass and luminosity functions. We checked that this correction does not vary by more than 1\% inside a bin.

\section{Correction of the main sequence at low-z}

\label{sect:lowzcorr}

When we use the \citet{Schreiber2015} analytic description of the evolution of the main sequence, the simulation exhibits an excess of SFRD at z$<0.5$ and overpredicts the bright end (S$_{100}>30$\,mJy) of the number counts at 100\,$\mu$m by $\sim$30\% (see Fig.\,\ref{fig:lowz}, red dashed line). We found a similar excess at the very bright end at other wavelengths. The simulation of \citet{Schreiber2016}, which uses the same description of the main sequence but different SEDs, has a similar excess at  S$_{100}$>30\,mJy (see their Fig.\,10). This is thus not a problem of SEDs. These bright sources have a median redshift of 0.22 and a median stellar mass of 2$\times$10$^{10}$\,M$_\odot$. At this stellar mass and redshift, the SFR$_{\rm MS}$ provided by \citet{Schreiber2015} is 0.1\,dex higher than \citet{Sargent2014} estimate based on a large compilation of data. We thus offset the sSFR$_{\rm MS}$ by $0.1 \times \frac{0.5-z}{0.5-0.22}$\,dex to correct for this offset. After implementing this correction, the simulation reproduces the SFRD and the bright end of the number counts (Fig.\,\ref{fig:lowz}, black solid line).

\section{The impact of clustering on stacking results as a function of redshift}

\label{app:stacking}


In Sect.\,\ref{sect:stacking}, we discussed the impact of clustering on the measurements by stacking. In this Appendix, we explain the technical details of our comparison between the corrections used by \citet{Schreiber2015} and \citet{Bethermin2015a} and the ones derived from our simulation.

\citet{Schreiber2015} used the real position of their stacked galaxies. They used their stellar mass and redshift to predict their mean expected flux densities at SPIRE wavelengths. They then built a map based on these predicted flux densities. They finally compared the mean flux density in their simulated catalog with the measurements by stacking in their simulated map. They used several methods to perform the photometry (small apertures, PSF-fitting photometry, signal in the central pixel). We chose to use the signal in the central pixel, since it is the easiest to simulate. After we had subtracted the mean value of the map to remove the background, we computed the mean flux density in all the SPIRE pixels hosting a galaxy in the input stacked catalog. The relative excess of flux density caused by clustered neighbors is computed using:
\begin{equation}
{\rm Relative \, \, excess} = \frac{S_{\rm stack} - \langle S_{\rm cat} \rangle}{\langle S_{\rm cat} \rangle},
\end{equation}
where $S_{\rm stack}$ is the mean flux density measured by stacking in the simulated map and $\langle S_{\rm cat} \rangle$ is the mean flux of the stacked sources in the simulated catalog. We found 13$\pm$1\,\%, 21$\pm$1\,\%, 34$\pm$1\,\% at 250, 350, and 500\,$\mu$m, respectively, which agrees well with the values of \citep{Schreiber2015} of $14_{-9}^{+14}$\,\%, $22_{-14}^{+19}$\,\%, and $39_{-23}^{+22}$\,\%, respectively.

\begin{figure}
\centering
\includegraphics{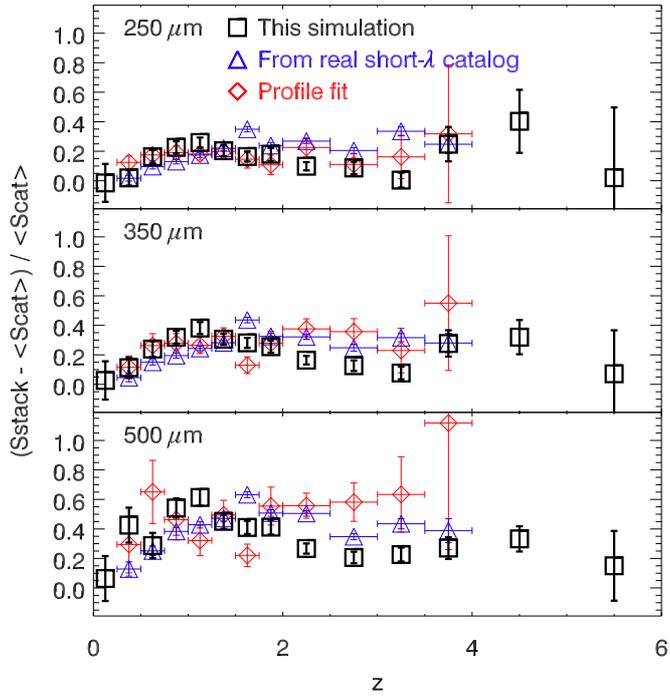}
\caption{\label{fig:stacking} Relative excess of flux density in stacking measurements caused by clustered neighbors as a function of redshift. These values were derived for a stellar mass selection (3$\times$10$^{10}$\,M$_\odot$). The black squares are the results from our simulation. The blue triangles were derived using a simulated map built from the real positions and the stellar masses of the galaxies in the COSMOS field \citep{Bethermin2015a}. The red diamonds were derived using a decomposition of the real stacked images into a point-like and an extended clustered component \citep{Bethermin2015a}.}
\end{figure}

We also compared our results with \citet{Bethermin2015b}, who used a redshift-dependent correction. To allow an easier comparison, we used the same stellar mass selection (3$\times$10$^{10}$\,M$_\odot$) and redshift bins and derived the stacking excess using the method described in the previous paragraph. Our results are presented in Fig.\,\ref{fig:stacking} (black squares). The relative excess caused by clustered neighbors is compatible with zero at z=0, rises up to a maximum at z$\sim$1, and slightly decreases with increasing redshift at z$>2$. This decrease of the relative excess at higher z was already discussed in \citet{Bethermin2015b} and was interpreted as the result of the rising of both the rarity and the infrared brightness of star forming galaxies with increasing redshift, causing a higher contrast between the massive galaxies and their environment. The first approach used in \citep{Bethermin2015b} uses an initial stacking to determine a mass-to-flux-density ratio evolving on the redshift (blue triangles). A simulated map is then produced from the real COSMOS catalog from the real position of the galaxies and their stellar mass. The relative excess caused by clustering is then estimated comparing the injected flux density  with the stacked flux density measured in this simulated map. This method neglects the diversity of the SEDs, the non-linearity of the M$_\star$-SFR relation, and the scatter around it. The trend between this method and our simulation agrees overall. However, our simulation found slightly lower values at z$>$2. This disagreement of $\sim$15\,\% is hard to explain and could come from the cosmic variance, a systematic effect at small scale in the real catalogs (incompleteness, problem of deblending), or a poor description of the small-scale clustering at z$>2$ in our simulation. The other method (red diamonds) is based on a stacking in map space and a fit of the resulting radial profile by both a point-like and an extended clustered component \citep{Heinis2013,Bethermin2015a,Welikala2016}. The results are similar to those obtained with the method described previously and our simulation at 250\,$\mu$m and 350\,$\mu$m. At 500\,$\mu$m, this method predicts higher values than the other ones at z$>2$. However, at 500\,$\mu$m, the decomposition is hard to perform, since the typical scale of the intra-halo clustering is close to the size of the beam.

Larger biases can be found if we stack fainter or more clustered galaxy populations. It can reach 90\% if we stack, for example, all sources with M$_\star > 10^{9}$\,M$_\sun$ at 500\,$\mu$m. The impact of clustering should thus be carefully checked while stacking \textit{Herschel} data. Our simulation will be particularly well suited to checking the accuracy of the stacking approaches in future studies.

\end{appendix}

\end{document}